\documentclass[journal]{IEEEtran}
\usepackage{amsmath,nccmath}
\usepackage{amssymb}
\usepackage{graphicx}
\usepackage{esint}
\usepackage{amsthm}
\usepackage{thmtools,thm-restate}
\usepackage{hyperref}
\usepackage{cleveref}
\usepackage{ifpdf}
\usepackage{cite}
\usepackage{color}
\usepackage{placeins}
\usepackage{wrapfig}
\usepackage{float}
\usepackage{tabularx}
\usepackage{colortbl}
\usepackage{pgfplots}
\usepgfplotslibrary{external}
\usepackage{tikz}
\usetikzlibrary{positioning,arrows.meta,fit,backgrounds,shapes.geometric}
\usepackage{tikzscale}
\pgfplotsset{compat=newest}
\usetikzlibrary{plotmarks}
\usetikzlibrary{arrows.meta}
\usepgfplotslibrary{patchplots}
\pgfplotsset{compat=newest}
\usetikzlibrary{plotmarks}
\usetikzlibrary{arrows.meta}
\usepgfplotslibrary{patchplots}
\usepackage{grffile}
\pgfplotsset{plot coordinates/math parser=false}
\newlength\figureheight
\newlength\figurewidth
\usepackage{pgfgantt}
\usepackage{pdflscape}
\usepackage[noend]{algorithmic}
\usepackage{algorithm}

\newtheorem{remark}{\textbf{Remark}}
\newtheorem{definition}{\textbf{Definition}}



\usepackage[
  bibbreaks=tight,
  paragraphs=tight,
  floats=tight,
  mathspacing=normal,
  wordspacing=tight,
  tracking=tight, 
  bibnotes=tight,
  charwidths=tight,
  mathdisplays=normal,
  leading=normal,
  indent=normal,
  lists=normal,
  bibliography=normal,
  title=normal,
  sections=normal,
  margins=normal
  ]{savetrees}

\title{\LARGE \bf Trust-Aware Sequential Decision Making and Rollout Planning for Resilient Multi-Robot Systems}
\author{Roee M. Francos*, Daniel Garces*, Orhan Eren Akgün, Nathaniel D. Bastian and Stephanie Gil
\thanks{This work is partially supported by AFOSR award $\#$FA9550-22-1-0223 and DARPA YFA award $\#$D24AP00319-00.}

\thanks{(*Co-primary authors) R.M.~Francos, D.~Garces, O.~Akgün, and S.~Gil are with the School of Engineering and Applied Sciences, Harvard University, Cambridge, MA 02138 USA, and N. Bastian is with the Whiting School of Engineering at the Johns Hopkins University, Baltimore, MD 21218 USA (e-mails: {\tt\small rfrancos@seas.harvard.edu, dgarces@g.harvard.edu, ndbastian@jhu.edu, sgil@seas.harvard.edu}).}
}
\begin{document}
\maketitle
\begin{abstract}
Sequential decision-making in multi-robot systems typically assumes that planning information is reliable and that agents execute the actions anticipated by the planner. Compromised agents can violate both assumptions, creating a mismatch between the planning model and physical execution. We study this problem in online multi-robot routing under localization spoofing. We introduce a distance-constrained spoofing model for monitor-aware adversaries, together with a tiered bipartite matching strategy that maximizes assignment influence while limiting spoofing magnitude. To mitigate such attacks, we develop a trust-aware monitor that combines probabilistic localization trust, calibrated using real GPS spoofing data, with behavioral evidence from task execution to classify agents and remove detected adversaries from subsequent planning. We further show that undetected adversaries can cause rollout to lose its expected cost-improvement behavior by violating planner-execution consistency. Trust-aware removal restores this consistency after detection, enabling stable routing and recovery of rollout's empirical advantage over the base policy. Experiments using real GPS spoofing datasets and San Francisco taxicab demand demonstrate effective detection and resilient routing across varying spoofing capabilities, adversarial fleet sizes, adaptive attacks, monitoring configurations, and rollout horizons.
\end{abstract}


%

\IEEEpeerreviewmaketitle

\section{Introduction}
Autonomous multi-robot systems increasingly rely on sequential decision-making algorithms to coordinate large teams in applications such as autonomous transportation, warehouse logistics, and aerial delivery \cite{brunke2022safe,bogyrbayeva2024machine,chib2023recent,verma2021multi,drew2021multi,dahiya2023survey}. These algorithms typically make decisions using reported agent states, task information, and predictions of future demand, while assuming that the information used for planning is reliable and that agents execute the prescribed actions. In practice, cyber attacks, localization spoofing, communication failures, or hardware faults can violate these assumptions \cite{sun2021survey,gil2023physicality,zardini2022analysis}. Unreliable agents can therefore do more than corrupt individual observations: they can create a mismatch between the system represented by the planner and the physical system that ultimately executes its decisions.

This mismatch is particularly consequential in online routing and task allocation, where decisions are repeatedly recomputed as tasks arrive and agent states evolve. A malicious or malfunctioning agent can attract assignments that it does not complete, inducing reassignment, wasting fleet capacity, delaying service, and increasing request cancellations \cite{wang2018ghost,nikitas2022deceitful,sun2021survey}. These effects can propagate over time and destabilize the routing process. Existing stability analyses for autonomous mobility systems generally assume fully cooperative fleets \cite{spieser2014,zhang2016control,garces2024approximate}; similarly, rollout-based planning relies on a predictive model that adequately represents the system executing the selected actions \cite{bertsekas2021multiagent,bertsekas2022lessons,bertsekas2023coursenew}. Compromised agents can violate both assumptions.

Localization spoofing provides a representative instance of this broader problem because routing decisions often depend directly on reported agent positions. A compromised agent can falsify its location to influence assignments while failing to provide the service anticipated by the routing policy. Moreover, a monitor-aware adversary need not spoof arbitrarily far: it may restrict its reported displacement to reduce detectability while retaining sufficient influence over routing decisions. This creates a trade-off between \emph{adversarial influence} and \emph{detectability} and motivates three challenges. First, the threat model must capture adversaries that strategically operate under spoofing constraints. Second, localization evidence alone may be insufficient against small or adaptive deviations, motivating complementary behavioral evidence from task execution. Third, trust estimates must be incorporated into the planning loop so that agents deemed unreliable no longer corrupt subsequent routing and lookahead decisions.

We address these challenges through a trust-aware sequential planning framework for resilient online multi-robot routing. Rather than treating trust only as a detection score, we use it to construct a \emph{trusted planning state}. At each decision epoch, the framework combines information-integrity observations with task-level behavioral evidence to classify active agents as \emph{cooperative}, \emph{suspect}, or \emph{adversarial}. Detected adversaries are removed from subsequent planning, and routing is performed over the resulting trusted \emph{active fleet}, defined as the set of agents that remain in the system following the monitor's decision at each time step. Execution outcomes then provide new evidence for future trust updates, forming a closed loop between trust estimation, planning, and execution.

We instantiate the framework for localization spoofing using two complementary trust signals. Localization trust probabilistically represents the reliability of reported positions and is calibrated using real GPS spoofing data, while behavioral trust captures task-level evidence such as successful pickups, assignment churn, and request expirations. We further introduce a distance-constrained adversarial spoofing model and tiered bipartite matching strategy in which monitor-aware adversaries coordinate their reported positions to influence routing assignments subject to bounded spoofing distances. This setting allows us to study attacks that remain disruptive while becoming increasingly difficult to distinguish from cooperative behavior using localization evidence alone.

A central focus of this work is the interaction between trust and rollout-based planning. Rollout evaluates candidate actions through lookahead simulations of the system dynamics \cite{bertsekas2021rollout,bertsekas2021multiagent}. If an undetected adversarial agent is treated as a cooperative resource during lookahead but does not execute the assumed service action, the simulated and physical system dynamics diverge. The resulting planner-execution mismatch can invalidate rollout's nominal cost-improvement behavior and cause it to perform worse than its base policy. Trust-aware removal reduces this mismatch by excluding detected adversaries from the active agent set used for planning, thereby restoring the consistency required for reliable rollout planning.

We evaluate the framework using real GPS spoofing datasets and a large-scale autonomous pickup-and-delivery simulation utilizing actual San Francisco taxicab demand data \cite{piorkowski2009crawdad}. The experiments consider varying spoofing ranges, adversarial fleet proportions, adaptive attacks, monitoring configurations, and rollout horizons. The results show that constrained adversaries can destabilize routing while remaining difficult to detect from localization information alone, that localization and behavioral evidence provide complementary detection capabilities, and that trust-aware adversarial removal can restore stable routing and rollout's empirical advantage over the IA-RA base policy.

The main contributions of this paper are:
\begin{itemize}
    \item \textbf{Monitor-aware adversarial spoofing:}
    We introduce a distance-constrained localization-spoofing model and tiered bipartite matching strategy that capture the trade-off between adversarial routing influence and spoofing detectability.
    
    \item \textbf{Trust-aware online routing and monitoring:}
    We develop a closed-loop framework that combines probabilistic localization trust with behavioral evidence to classify agents and construct a trusted active fleet used for subsequent planning steps.
    
    \item \textbf{Trust-aware rollout under planner-execution mismatch:}
    We identify planner-execution consistency as a key requirement for rollout-based routing, demonstrate how undetected adversaries violate this condition, and show how trust-aware removal enables rollout to recover stable operation and its empirical advantage over the IA-RA base policy.
    
    \item \textbf{Empirical validation:}
    Using real GPS spoofing data and San Francisco mobility demand, we characterize localization-based trust and evaluate the proposed framework across diverse adversarial, monitoring, and rollout configurations, including adaptive and distance-constrained attacks.
\end{itemize}

The remainder of this paper is organized as follows. Section~\ref{sec:related_work} reviews related work. Section~\ref{sec:problem_formulation} formulates the adversarial online-routing problem and defines the true and perceived system states, service model, and performance criteria. Section~\ref{sec:trust_aware_framework} presents the trust-aware sequential planning framework and its closed-loop information flow. Section~\ref{sec:adversarial_spoofing} develops the monitor-aware adversarial spoofing model and strategy. Section~\ref{sec:trust_monitor} introduces the localization- and behavioral-trust mechanisms, agent classification, and trusted-fleet update. Section~\ref{sec:trust_aware_rollout} analyzes planner-execution mismatch and presents trust-aware rollout. Section~\ref{sec:experiments} presents the experimental evaluation, and Section~\ref{sec:discussion} discusses the broader implications and limitations of the framework. Finally, Section~\ref{sec:conclusion} provides a summary of the key contributions of the paper to conclude.

\section{Related Work}
\label{sec:related_work}

This work lies at the intersection of adversarial multi-robot routing, localization spoofing and trust estimation, rollout-based sequential decision making, and resilient multi-robot systems. We organize the related work according to the progression of the problem studied in this paper. We first review how non-cooperative or compromised agents affect online routing and task allocation, with emphasis on attacks that manipulate the information used for assignment decisions. We then discuss localization spoofing and localization-integrity estimation as mechanisms for corrupting and assessing planning information. Next, we review rollout-based sequential decision making and the dependence of model-based planning on consistency between predicted and executed system behavior. Finally, we position the proposed approach relative to broader work on resilient monitoring and multi-robot coordination.

\subsection{Adversarial Multi-Robot Routing and Task Allocation}

Autonomous transportation and multi-robot systems rely on routing and task-allocation algorithms to repeatedly assign agents to spatially distributed requests \cite{chakraaoptimization2023}. Most existing methods assume cooperative agents and reliable state information, assumptions that simplify policy design and stability analysis but may fail when agents are selfish, malfunctioning, or compromised.

Resilient task allocation has been studied through adaptive control \cite{emam2021adaptive} and resilient assignment strategies for heterogeneous multi-robot teams \cite{mayya2021resilient}. These approaches improve robustness to environmental uncertainty, robot failures, and degraded execution, but do not directly address adversaries that manipulate the information used to compute task assignments.

Recent work has also examined the effect of non-cooperative agents on routing stability. In \cite{francos2025provably}, adversarial or malfunctioning agents are modeled through bounded service delays during pickup and delivery, showing that even limited deviations from cooperative behavior can cause outstanding requests to accumulate and destabilize the system. This model captures failures during task execution but does not consider adversaries that manipulate the state information used to determine assignments.

Information manipulation creates a different failure mode because adversaries can influence the decision process before execution occurs. In online routing, incorrect state reports can alter request assignments to agents, with their effects propagating through repeated reassignment, request expiration, and growing outstanding demand. Prior work has shown that localization spoofing can destabilize reassignment-based policies such as Instantaneous Assignment with Reassignment (IA-RA) \cite{francosgarcespolicy2026}. However, existing routing attack models do not capture monitor-aware adversaries that deliberately restrict their spoofing magnitude to reduce detectability while retaining influence over assignments. This motivates adversarial models that explicitly account for the trade-off between routing influence and observable localization deviation.

\subsection{Localization Spoofing and Trust Estimation}

Localization spoofing poses a significant cyber-physical vulnerability because routing, planning, and control algorithms often depend directly on reported position information. GPS spoofing have been successfully carried out against autonomous vehicles and UAVs, including complete vehicle takeover \cite{sathaye2022experimental}, stealthy drift attacks \cite{dasgupta2024unveiling} and attacks against multi-sensor localization systems \cite{shen2020drift}. Related work has considered spoofing of LiDAR-based localization and SLAM \cite{suzuki2025lab,fukunaga2024random,nagata2025slamspoof}, as well as attacks on cooperative perception systems \cite{li2024advgps}. A broader overview of sensor-spoofing threats in autonomous robotic systems is provided in \cite{xu2023sok}.

A complementary line of work focuses on detecting spoofing and estimating localization integrity using statistical inference, sensor fusion, and learning-based techniques \cite{alhoraibi2024detection,badar2025deepspoofnet}. These methods provide mechanisms for determining whether localization information is likely to be reliable. However, localization-integrity estimation and downstream planning are typically treated as separate problems: a detector produces an integrity estimate or attack classification, while the routing or planning algorithm continues to be analyzed independently.

This separation becomes important when attacks are difficult to identify from localization observations alone. A monitor-aware adversary may reduce the magnitude of its spoofing so that its localization signals remain closer to nominal behavior while still influencing task allocation. In such settings, localization evidence can be complemented by behavioral evidence obtained from downstream execution outcomes. The problem considered in this study therefore requires not only estimating the reliability of localization information, but also combining that estimate with evidence of how an agent affects the task-allocation process and using the resulting trust assessment to modify subsequent planning.

\subsection{Online Routing and Rollout-Based Sequential Decision Making}

Online routing is challenging because requests arrive dynamically, agent states evolve over time, and the joint state-action space grows rapidly with the number of agents and tasks. Exact dynamic programming is generally impractical at realistic scales, motivating approximate assignment and planning approaches. Existing methods include instantaneous assignment algorithms \cite{bertsekas1979distributed,duan2014linear,bertsimas2019online}, local-search heuristics such as 2-opt \cite{croes1958method,yannakakis1990analysis}, stochastic optimization \cite{lowalekar2018online}, and reinforcement-learning approaches including approximate value iteration and deep reinforcement learning \cite{ulmer2019offline,farazi2021deep,ahamed2021deep}.

Rollout provides a model-based reinforcement learning approach for improving a computationally tractable base policy through finite-horizon lookahead simulation without solving the underlying stochastic dynamic program exactly \cite{bertsekas2021rollout,bertsekas2021multiagent}. Its cost improvement property relies on the predictive model used during future lookahead being representative of the system that executes the selected actions. Under nominal operation, this assumption is natural when all agents follow the routing policy used by the planner.

Compromised agents create a different setting. If an adversarial agent is represented during lookahead as a cooperative resource but does not execute the service behavior assumed by the planner, the simulated and physical system dynamics diverge. The resulting cost-to-go estimates may no longer accurately rank candidate actions, and the nominal improvement behavior of rollout need not be preserved. Existing rollout formulations generally do not consider adversarial corruption of the agents or information represented in the planning model. This motivates studying trust not only as a detection mechanism, but also as a means of maintaining consistency between the set of agents used for planning and the system that executes the resulting decisions.

\subsection{Resilient Multi-Robot Systems and Trust-Aware Monitoring}

The vulnerability of learning-based and sequential decision-making systems to uncertainty and adversarial manipulation has attracted increasing attention. Prior work studied adversarial observation attacks \cite{gleave2020adversarial}, learned disruption policies \cite{mo2022attacking}, resilient routing under uncertainty \cite{shi2023robust}, and denial-of-service attacks in mobility systems \cite{thai2016resiliency}. Explainability and saliency-based techniques have also been proposed for attack detection \cite{wang2024explainable,hickling2023robust}, although their reliability may degrade under complex or adaptive adversarial behavior \cite{behzadan2017whatever,zhang2020robust}.

More broadly, resilient multi-robot systems have used redundancy, fault tolerance, and reconfiguration to maintain team performance under failures or attacks \cite{prorok2021beyond,zhou2021multi}. Representative threats include identity spoofing \cite{ballotta2024role}, dissemination of false information \cite{cavorsi2024exploiting}, and coordinated adversarial behavior in multi-robot teams \cite{francos2023role,francos2024defense}. These works provide important mechanisms for resilient coordination, but adversarial modeling, information-integrity estimation, behavioral monitoring, and sequential planning are often considered separately.

The present work connects these components in the setting of online multi-robot routing. In contrast to approaches that use trust primarily to detect corrupted information, we study how information-level and behavioral evidence can be used to determine which agents should remain part of the planning set. We further examine the implications of this decision for model-based sequential planning, focusing on the planner-execution mismatch created by undetected adversaries and the role of trust-aware removal in recovering reliable rollout behavior.

\section{Problem Formulation}
\label{sec:problem_formulation}

We consider online multi-agent task allocation and routing in the presence of unreliable localization information and non-cooperative agent behavior. The central feature of the formulation is the distinction between the information available to the routing policy and the state that governs physical execution. The routing server computes assignments from a \emph{perceived state}, which may contain corrupted localization reports, whereas service outcomes and system evolution are determined by the \emph{true state} and by the actions actually executed by the agents. This distinction provides the basis for the planner-execution mismatch studied later in the paper.

\subsection{System Model and Information Structure}
We consider an autonomous pickup-and-delivery system operating on a strongly connected directed graph
$\mathcal{G}=(\mathbb{V},\mathbb{E})$.
Let $d_{\mathcal{G}}(u,v)$ denote the shortest-path travel time from node $u\in\mathbb{V}$ to node $v\in\mathbb{V}$, and define the graph diameter as,
\begin{equation}
D(\mathcal{G})=
\max_{u,v\in\mathbb{V}}
d_{\mathcal{G}}(u,v)
\end{equation}
A fleet of autonomous agents services transportation requests that arrive stochastically over time. Let $\mathcal{L}_0$ denote the initial fleet, consisting of cooperative agents $\mathcal{C}_0$ and adversarial agents $\mathcal{A}_0$, with 
\[
\mathcal{L}_0=\mathcal{C}_0\cup\mathcal{A}_0,
\qquad
\mathcal{C}_0\cap\mathcal{A}_0=\emptyset
\]
The initial fleet size is $N_0=|\mathcal{L}_0|$, and the initial adversarial fraction is denoted by,
\[
F_0=\frac{|\mathcal{A}_0|}{N_0}
\]
At time $t$, let $\mathcal{L}_t\subseteq\mathcal{L}_0$ denote the set of agents currently participating in routing, with $N_t=|\mathcal{L}_t|$. The corresponding cooperative and adversarial subsets are
$\mathcal{C}_t=\mathcal{C}_0\cap\mathcal{L}_t$ and
$\mathcal{A}_t=\mathcal{A}_0\cap\mathcal{L}_t$.
The approach describing how trust information is used to modify the set of agents considered for planning is introduced separately in Section~\ref{sec:trust_aware_framework}.

Let $\bar{\mathcal{R}}_t$ denote the set of outstanding requests at time $t$. We distinguish between two system representations. The \emph{true state},
\begin{equation}
x_t=
\left(
\boldsymbol{\nu}_t,
\boldsymbol{\tau}_t,
\bar{\mathcal{R}}_t
\right)
\end{equation}
contains the physical agent states, where
$\boldsymbol{\nu}_t=(\nu_t^i)_{i\in\mathcal{L}_t}$
and
$\boldsymbol{\tau}_t=(\tau_t^i)_{i\in\mathcal{L}_t}$
denote the true locations and remaining trip times of the participating agents. The centralized routing server instead observes the \emph{perceived state},
\begin{equation}
\hat{x}_t=
\left(
\hat{\boldsymbol{\nu}}_t,
\hat{\boldsymbol{\tau}}_t,
\bar{\mathcal{R}}_t
\right)
\end{equation}
where
$\hat{\boldsymbol{\nu}}_t=(\hat{\nu}_t^i)_{i\in\mathcal{L}_t}$
and
$\hat{\boldsymbol{\tau}}_t=(\hat{\tau}_t^i)_{i\in\mathcal{L}_t}$
denote the localization and trip-time information available to the server. Cooperative agents report their localization information truthfully, so that
$\hat{\nu}_t^i=\nu_t^i$ for $i\in\mathcal{C}_t$, whereas adversarial agents may report locations satisfying
$\hat{\nu}_t^a\neq\nu_t^a$.
Consequently, the routing policy generally operates on $\hat{x}_t$, while physical execution is governed by $x_t$.

\subsection{Request Arrival and Service Model}
A transportation request is represented as,
\[
r=\langle \rho_r,\delta_r,t_r,\phi_r\rangle
\]
where $\rho_r\in\mathbb{V}$ and $\delta_r\in\mathbb{V}$ denote its pickup and drop-off locations, $t_r$ is its arrival time, and $\phi_r$ indicates whether the request has been picked up.

At each time step, the number of new requests is sampled from a distribution $p_{\eta}$. Conditional on a request arrival, its pickup and drop-off locations are sampled from distributions $p_{\rho}$ and $p_{\delta}$, respectively. These distributions are estimated from historical demand data. We assume that finitely many requests arrive at each time step and that request locations are sampled independently across arrivals.

A request remains in $\bar{\mathcal{R}}_t$ until it is either picked up or canceled. If it is not picked up within a prescribed waiting-time limit, it expires and is counted as canceled. Thus, the outstanding-request set evolves through new arrivals, successful pickups, and expirations.

\subsection{Adversarial Agent Model}
We consider adversarial agents that manipulate the information used by the routing policy and may deviate from the service behavior assumed by the centralized planner. The reported location $\hat{\nu}_t^a$ of an adversarial agent $a\in\mathcal{A}_t$ may therefore differ from its true location $\nu_t^a$.

\begin{definition}[Adversarial Agent Localization Spoofing Model]
\label{def:adversarial_agent}
An agent $a\in\mathcal{A}_0$ follows the adversarial localization spoofing model if it satisfies the following:
\begin{enumerate}
\item It may manipulate its reported location prior to a routing decision.
\item It is associated with a unique immutable identity and cannot create Sybil identities.
\item It may disregard assigned service actions and, in the attack setting considered here, does not service requests assigned to it.
\item It has knowledge of the routing policy and assignment mechanism.
\item It may coordinate its reported localization information with other adversarial agents.
\end{enumerate}
\end{definition}

Definition~\ref{def:adversarial_agent} specifies the common capabilities assumed throughout the paper without prescribing how an adversary selects its spoofed location. Section~\ref{sec:adversarial_spoofing} subsequently introduces the monitor-aware distance-constrained threat model and the corresponding coordinated spoofing strategy.

\subsection{Routing Decisions and Physical Execution}
At each decision epoch, the centralized server applies a routing policy $\pi_t$ to the perceived state and computes a commanded joint action,
\begin{equation}
u_t=\pi_t(\hat{x}_t)
\label{eq:commanded_action}
\end{equation}
We distinguish this commanded action from the action that is actually executed by the fleet. Let $u_t^{\mathrm{exec}}=
\left(u_t^{i,\mathrm{exec}}\right)_{i\in\mathcal{L}_t}$ denote the joint executed action. Cooperative agents follow the routing command, whereas an adversarial agent may satisfy $u_t^{a,\mathrm{exec}}\neq u_t^a$.
The true system therefore evolves according to,
\begin{equation}
x_{t+1} =
f\left(
x_t,
u_t^{\mathrm{exec}},
\eta_t,
\boldsymbol{\rho}_t,
\boldsymbol{\delta}_t
\right)
\label{eq:true_system_dynamics}
\end{equation}
where $\eta_t$ denotes the number of new requests with their pickup and drop-off locations collected in
$\boldsymbol{\rho}_t$ and $\boldsymbol{\delta}_t$, respectively.

Equation~\eqref{eq:true_system_dynamics} emphasizes two distinct sources of discrepancy between planning and execution. First, the routing policy computes $u_t$ using the potentially corrupted perceived state $\hat{x}_t$. Second, even after an action is selected, adversarial agents may not execute the service behavior assumed by the routing policy. These discrepancies are central to the trust-aware planning problem considered in this paper.

\subsection{Performance and Stability Criterion}
We evaluate routing performance through outstanding demand and request cancellations. Let
$|\bar{\mathcal{R}}_t|$
denote the number of requests awaiting pickup at time $t$, and let
$|\mathcal{R}^{\mathrm{can}}_{0:t}|$
denote the cumulative number of requests that have expired up to time $t$. We define the realized stage cost as,
\begin{equation}
g_t(\hat{x}_t, u_t^{\mathrm{exec}}, \eta_t, \boldsymbol{\rho}_t, \boldsymbol{\delta}_t) = 
|\bar{\mathcal{R}}_t|
+
|\mathcal{R}^{\mathrm{can}}_{0:t}|
\label{eq:stage_cost}
\end{equation}
The first term measures the current service backlog, while the second captures accumulated service failures. Because canceled requests remain in the cumulative term, persistent request expiration results in increasing long-term cost.

For a routing policy $\pi$, with commanded actions generated according to
$u_k=\pi_k(\hat{x}_k)$, define the average cost,
\begin{equation}
J\pi^{\mathrm{avg}}(\hat{x}_t)
=
\limsup_{T\rightarrow\infty}
\frac{1}{T}
\mathbb{E}
\left[
\sum_{k=t}^{T} g_k(\hat{x}_k, u_k^{\mathrm{exec}}, \eta_k, \boldsymbol{\rho}_k, \boldsymbol{\delta}_k)
\right]
\label{eq:average_cost}
\end{equation}

\begin{definition}[Average Cost Stability]
\label{def:routing_stability_cost_of_policy_average_cost}
A policy $\pi$ is stable if $J_\pi^{\mathrm{avg}}(\hat{x}_t)<\infty$, for all observed states $\hat{x}_t$ for all $t$.
\end{definition}

Under this definition, stability requires the routing system to prevent persistent accumulation of outstanding and canceled requests. Conversely, unbounded growth in these quantities indicates that service demand is accumulating faster than the fleet can process it, consistent with the average-cost stability criterion used in \cite{francosgarcespolicy2026}.

\subsection{Planner-Execution Mismatch}
The formulation above exposes the fundamental challenge considered in this work. The routing policy selects actions from the perceived state $\hat{x}_t$, but realized performance depends on the true state $x_t$ and on the executed action $u_t^{\mathrm{exec}}$. Under cooperative operation, these representations are consistent: agents report reliable state information and execute the prescribed service actions. Adversarial localization spoofing can break both relationships by corrupting the state used to compute assignments and by causing an agent's executed behavior to differ from that anticipated by the planner.

We refer to this discrepancy as \emph{planner-execution mismatch}. In instantaneous routing, such mismatch can result in inappropriate assignments, reassignment churn, and service failures. In model-based lookahead methods such as rollout, it additionally causes simulated trajectories to represent behavior that may not occur during physical execution. Our objective is therefore to use observable reliability evidence to construct a planning state that excludes sufficiently unreliable agents, thereby reducing planner-execution mismatch while retaining as much cooperative fleet capacity as possible. The next section introduces the trust-aware sequential planning framework used to achieve this objective.

\section{Trust-Aware Sequential Planning Framework}
\label{sec:trust_aware_framework}
\begin{figure*}
\centering
\includegraphics[width=0.99\linewidth]{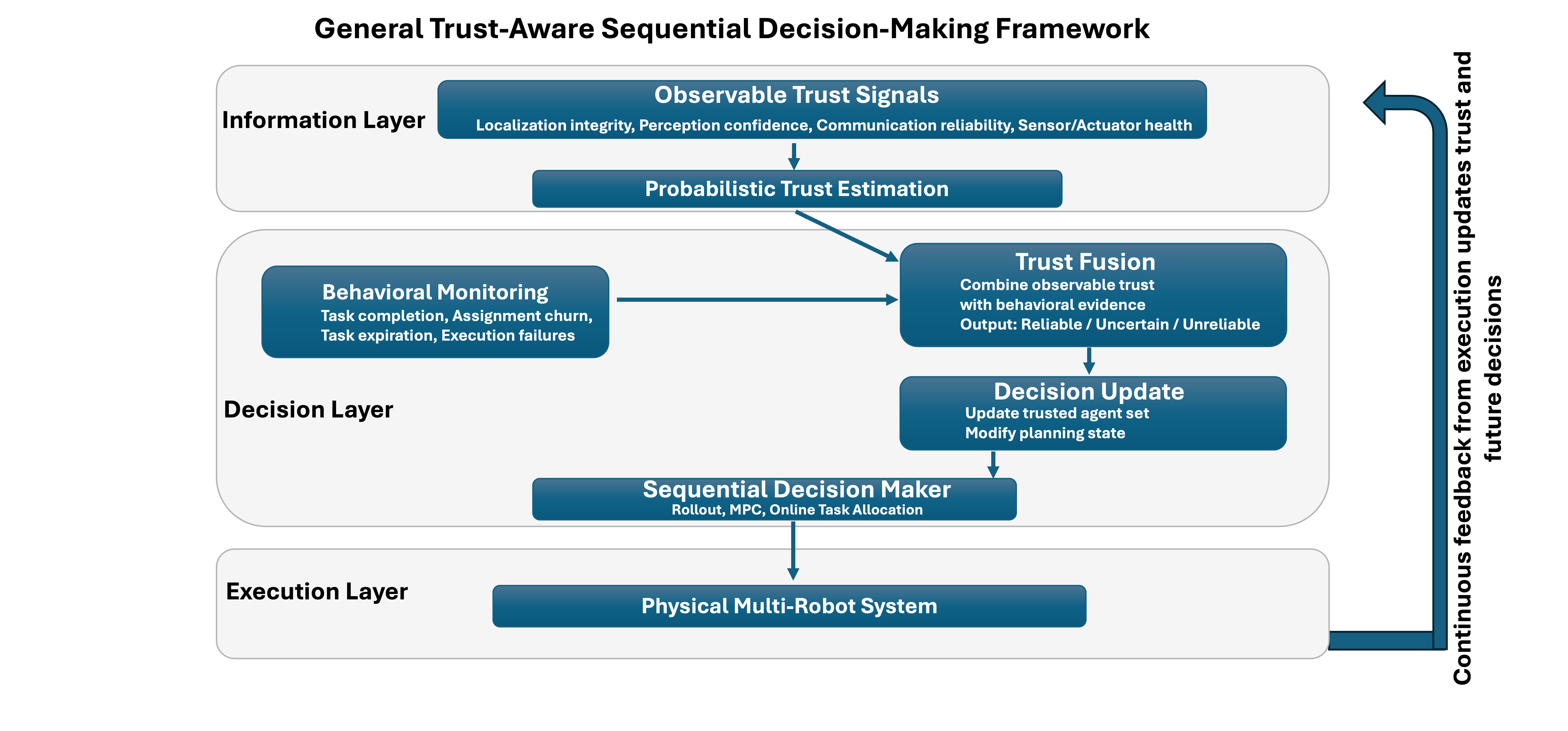}
    \vspace{-10pt}
\caption{Trust-aware sequential planning framework. Information-integrity observations and behavioral evidence are used to update agent trust classifications and construct a trusted planning state. The planner computes actions using this filtered state, while execution generates new observations and task-level outcomes that are incorporated into subsequent trust updates.}
\label{fig:general_framework}
\end{figure*}

The problem formulation in Section~\ref{sec:problem_formulation} distinguishes between the perceived state used to compute routing decisions and the true system that executes those decisions. When the information or behavior associated with an agent is unreliable, retaining that agent in the planning representation can create planner-execution mismatch. We therefore introduce a trust-aware sequential planning framework whose purpose is to determine, from observable evidence, which information and agents should remain part of the state used for subsequent planning. The framework separates trust-aware sequential planning into three interacting functions: information assessment, trust-aware state construction and planning, and execution. These components form the closed-loop architecture illustrated in Fig.~\ref{fig:general_framework}, in which observations are transformed into reliability evidence, reliability evidence modifies the planning state, and execution generates new evidence for subsequent decision epochs.

The central idea is to use trust not only as an estimate of information reliability, but as a mechanism for constructing a \emph{trusted planning state}. At each decision epoch, the framework evaluates reliability evidence associated with the participating agents, updates their trust classifications, filters the perceived state according to these classifications, and applies the underlying sequential decision-making policy to the resulting state. Execution then generates new information-integrity observations and task-level outcomes that are used to refine trust at subsequent decision epochs. 

\subsection{Framework Overview}
\label{subsec:framework_overview}
The framework separates trust-aware sequential planning into three interacting functions: \emph{information assessment}, \emph{trust-aware state construction and planning}, and \emph{execution}.

First, information assessment produces reliability evidence from observable signals associated with each agent. These observations may characterize the integrity of the information supplied to the planner, such as localization or perception confidence. Because an agent may provide information that appears plausible while nevertheless behaving inconsistently with the planner's assumptions, information-level evidence is complemented by behavioral evidence obtained from execution outcomes.

Second, the trust-aware decision process combines the available evidence to maintain an agent-level reliability classification. Let
\[
\chi_i(t)\in\{\mathsf{C},\mathsf{S},\mathsf{A}\}
\]
denote the trust state of agent $i$ at decision epoch $t$, corresponding to \emph{cooperative}, \emph{suspect}, and \emph{adversarial}, respectively. The classification determines whether the agent remains available to the planner. Agents for which the accumulated evidence is insufficient or ambiguous remain in the suspect state rather than being immediately removed. This allows the framework to defer consequential decisions until stronger evidence is available.

Finally, the sequential planner operates on the state retained after this trust-aware filtering step. The framework does not prescribe a particular planning algorithm: the same interface separates reliability assessment from the policy used to compute actions. The resulting command is executed by the physical system, and the observed execution outcomes provide new evidence for subsequent trust updates. In this way, trust estimation and planning are coupled through a feedback loop rather than treated as independent detection and control problems.

\subsection{Trusted Planning State}
\label{subsec:trusted_planning_state}

Let $\hat{x}_t$ denote the perceived state available before trust-based filtering, as defined in Section~\ref{sec:problem_formulation}, and let
\[
\boldsymbol{\chi}_t = 
\bigl(\chi_i(t)\bigr)_{i\in\mathcal{L}_t}
\]
collect the current trust classifications. We define a trust-aware filtering operator $\mathcal{F}$ that constructs the state used for planning:
\begin{equation}
\tilde{x}_t =
\mathcal{F}\left(\hat{x}_t,\boldsymbol{\chi}_t\right)
\label{eq:trusted_planning_state}
\end{equation}
The resulting $\tilde{x}_t$ is referred to as the \emph{trusted planning state}. The trust-aware policy then computes
\begin{equation}
u_t = 
\pi_t \left(\tilde{x}_t\right)
\label{eq:trust_aware_policy}
\end{equation}
rather than applying the policy directly to the unfiltered perceived state $\hat{x}_t$.

The term \emph{trusted} does not imply that every element retained in $\tilde{x}_t$ is known to be reliable. In particular, agents classified as suspect may remain available to the planner while additional evidence is accumulated. Instead, $\tilde{x}_t$ represents the planning state after agents for which sufficient evidence of adversarial behavior has been obtained are excluded. This distinction is important because aggressive removal can reduce useful system capacity, whereas delayed removal allows unreliable agents to continue influencing decisions.

The purpose of this filtering operation is therefore to reduce planner-execution mismatch while preserving as much reliable planning capacity as possible. If unreliable agents remain undetected, $\tilde{x}_t$ may still contain information or resources whose actual execution differs from the behavior assumed by the planner. As evidence accumulates and such agents are identified, the planning representation becomes progressively better aligned with the set of agents available for reliable execution.

\subsection{Closed-Loop Decision Epoch}
\label{subsec:closed_loop_decision_epoch}
To make the information flow explicit, one trust-aware decision epoch proceeds as follows.

At the beginning of epoch $t$, the system receives the current reported state information together with any newly available information-integrity observations. Behavioral evidence generated by previous execution, such as task completion or failure, is also available to the trust mechanism. These observations are used to update the trust state $\boldsymbol{\chi}_t$.

The framework then applies the filtering operation in~\eqref{eq:trusted_planning_state} to construct $\tilde{x}_t$. The underlying planning policy uses this state to compute the commanded action according to~\eqref{eq:trust_aware_policy}. The physical system executes the resulting command; as established in Section~\ref{sec:problem_formulation}, the executed action may differ from the commanded action when unreliable agents remain among those considered during planning. Execution updates the physical state and generates new information and behavioral outcomes, which become evidence for future trust updates. Thus, the closed-loop sequence is,
\begin{align*}
    & \text{observations}
    \;\rightarrow\;
    \text{trust update}
    \;\rightarrow\;
    \text{trusted planning state} \\
    & \qquad
    \;\rightarrow\;
    \text{planning}
    \;\rightarrow\;
    \text{execution}
    \;\rightarrow\;
    \text{new observations}
\end{align*}
This ordering also separates the information available at planning time from evidence that becomes available only after execution. Trust-aware planning therefore does not require knowledge of an agent's true type; it operates from observable reliability evidence and modifies the planning representation as that evidence accumulates.

\subsection{Instantiation for Resilient Online Routing}
\label{subsec:framework_routing_instantiation}
The remainder of the paper instantiates this architecture for the online routing problem defined in Section~\ref{sec:problem_formulation}. In this setting, the information whose reliability must be assessed is the reported localization state. The information-level component therefore uses probabilistic localization-integrity evidence, while the behavioral component evaluates the consequences of an agent's participation in task allocation through request pickups, assignment churn, and request expirations.

For the routing instantiation, the trust-aware filtering operation is implemented by updating the active planning fleet. After classification at epoch $t$, define
\begin{equation}
\mathcal{L}_{t^+} = 
\mathcal{L}_t
\setminus
\left\{
i\in\mathcal{L}_t:
\chi_i(t)=\mathsf{A}
\right\}
\label{eq:framework_trusted_active_fleet}
\end{equation}
where $t^+$ denotes the instant after trust-based enforcement and before the subsequent routing decision. The trusted planning state is then the perceived state restricted to this fleet, $\hat{x}_{t^+} = 
\hat{x}_t\big|_{\mathcal{L}_{t^+}}$. 
Thus, $\hat{x}_{t^+}$ is the routing-specific realization of the general trusted planning state $\tilde{x}_t$ in~\eqref{eq:trusted_planning_state}.

The detailed components that produce this update are developed in the following sections. Section~\ref{sec:adversarial_spoofing} introduces the monitor-aware localization-spoofing model used to evaluate the framework under adversarial conditions. Section~\ref{sec:trust_monitor} develops the localization and behavioral-trust mechanisms, their fusion into $\chi_i(t)$, and the enforcement rule that constructs $\mathcal{L}_{t^+}$. Section~\ref{sec:trust_aware_rollout} then studies how planning over the filtered state affects rollout-based routing when adversarial agents can otherwise create a mismatch between simulated and executed behavior.

The resulting architecture should therefore be viewed as a coupling between \emph{reliability assessment} and \emph{sequential planning}, rather than as a standalone localization-spoofing detector. In the routing instantiation, adversarial removal is the mechanism used to modify the planning state. The subsequent sections specify how the evidence required for that decision is generated and how the resulting state is used by the routing planner.

\section{Monitor-Aware Adversarial Spoofing}
\label{sec:adversarial_spoofing}

The trust-aware framework in Section~\ref{sec:trust_aware_framework} is intended to reduce planner-execution mismatch caused by unreliable information and agent behavior. To evaluate this framework under strategic interference, we now specify how adversarial agents select the localization reports presented to the routing system. This section therefore defines the \emph{threat model used to stress-test the proposed framework} by deliberately exposing it to challenging adversarial behaviors designed to increase planner-execution mismatch; it is not a component of the trust-aware planning architecture itself.

The adversarial agent capabilities are defined in Section~\ref{sec:problem_formulation}. Here, we additionally consider \emph{monitor-aware} adversaries that recognize that large localization discrepancies may provide stronger evidence of manipulation. Rather than allowing arbitrary spoofed positions, these adversaries restrict the magnitude of their localization deviations while coordinating their reports to retain influence over routing decisions. This creates the influence-detectability trade-off studied throughout the remainder of the paper.

\subsection{Monitor-Aware Spoofing Model and Decision-Epoch Timing}
\label{subsec:monitor_aware_threat_model}
Consider an active adversarial agent $a\in\mathcal{A}_t$ with true location $\nu_t^a$. Under \emph{unconstrained localization spoofing}, the agent may report any position in $\mathbb{V}$. We instead restrict its reported position to the graph neighborhood, $\mathcal{B}_t^a(d_{\max})
=
\left\{
v\in\mathbb{V}:
d_{\mathcal{G}}(\nu_t^a,v)\leq d_{\max}
\right\}$, 
where $d_{\max}\geq 0$ is the maximum allowable spoofing distance.

\begin{definition}[Monitor-Aware Distance-Constrained Spoofing]
\label{def:monitor_aware_spoofing}
An adversarial agent $a\in\mathcal{A}_t$ follows the
\emph{monitor-aware distance-constrained spoofing model} if it satisfies the
adversarial capabilities of Definition~\ref{def:adversarial_agent} and selects
its reported position such that,
\begin{equation}
\hat{\nu}_t^a\in\mathcal{B}_t^a(d_{\max}),
\qquad\text{or equivalently}\qquad
d_{\mathcal{G}}(\nu_t^a,\hat{\nu}_t^a)\leq d_{\max}
\label{eq:distance_constrained_spoofing}
\end{equation}
\end{definition}
The parameter $d_{\max}$ controls the adversary's ability to trade spatial influence for localization discrepancy. Small values constrain the attacker to reports close to its true position, potentially reducing the evidence available to a localization-integrity monitor, whereas larger values provide access to a greater set of spoofed locations. When $d_{\max}\geq D(\mathcal{G})$, the constraint is inactive and the model reduces to unconstrained spoofing.

The term \emph{monitor-aware} refers to the adversary's recognition that spoofing magnitude affects detectability. The attack considered here does not require direct access to the internal trust state or classification thresholds of the monitor. 

\begin{figure}[t]
\centering
\includegraphics[
    width=0.95\linewidth,
    height=0.4\textheight,
    keepaspectratio
]{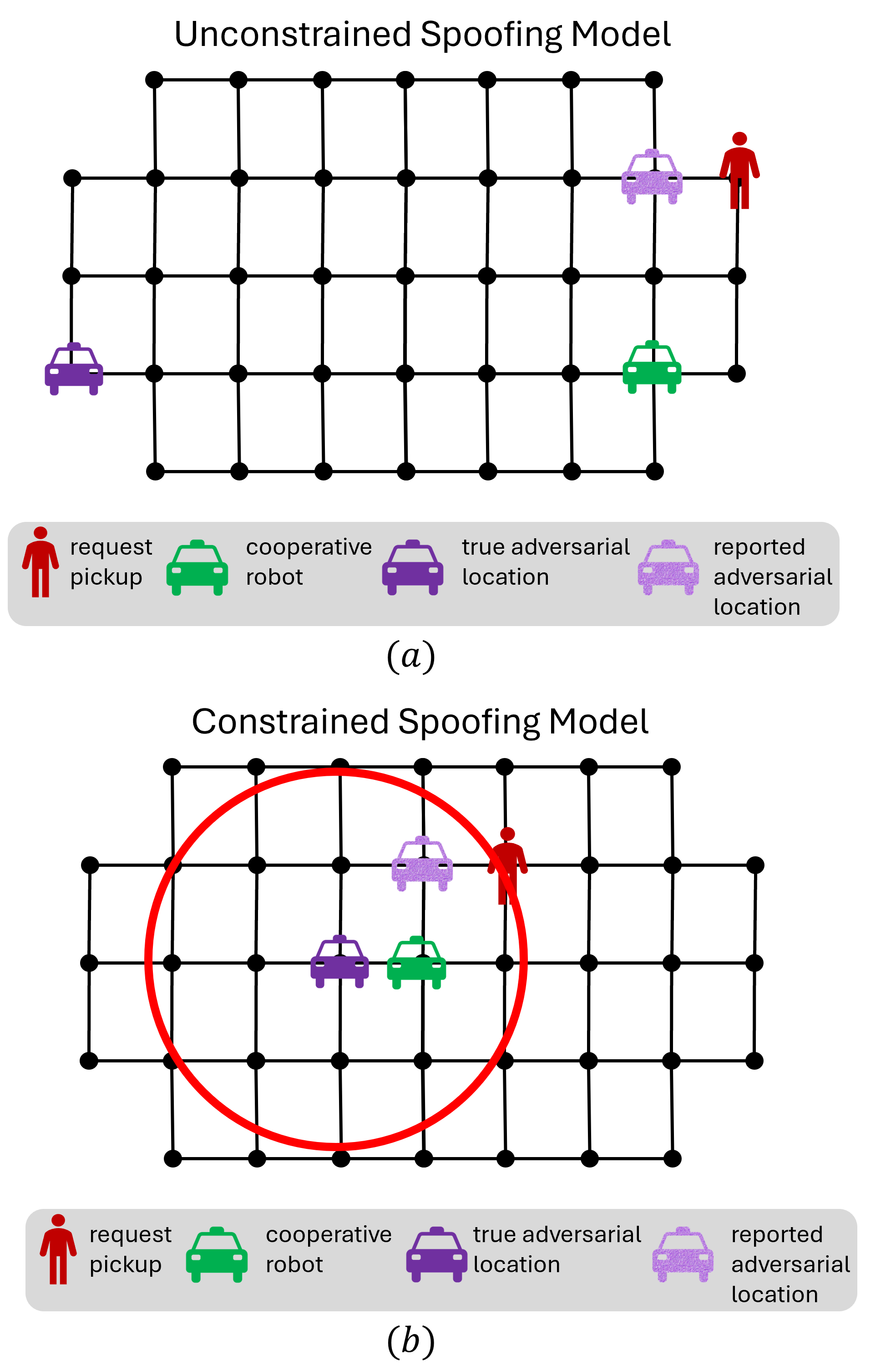}
\vspace{-7pt}
\caption{Localization-spoofing threat models. Unconstrained spoofing permits an adversarial agent to report any location on the graph, whereas monitor-aware distance-constrained spoofing restricts the reported location to lie within graph distance $d_{\max}$ of the true position.}
\label{fig:unconstrained_and_constrained_spoofing_model}
\vspace{-7pt}
\end{figure}

The timing of the attack is consistent with the trust-aware decision epoch introduced in Section~\ref{subsec:closed_loop_decision_epoch}. At the beginning of epoch $t$, active adversaries select their reported locations $\{\hat{\nu}_t^a\}_{a\in\mathcal{A}_t}$ before the trust update and routing decision. These reports enter the perceived state $\hat{x}_t$ and generate the localization-integrity evidence available to the monitor. The monitor then updates agent classifications and constructs the trusted planning state before the routing policy is applied. Consequently, an adversarial report can influence the current routing decision only if the corresponding agent remains among those considered for planning after monitoring.

\subsection{Cooperative-Feasible Adversarial Targets}
\label{subsec:adversarial_target_set}
We next define the coordinated strategy used by adversarial agents to select spoofed locations. The attack is constructed as a white-box stress test: the adversarial team is assumed to know the current outstanding requests, their remaining lifetimes, the routing mechanism, the cooperative/adversarial fleet membership, and the current locations used to construct the attack. These assumptions provide the adversary with favorable information and allow us to evaluate the trust-aware framework under coordinated interference.

Let $\bar{\mathcal{R}}_t$ denote the set of outstanding requests, and let
$T_r(t)$ denote the remaining time before request $r$ expires. The adversarial team targets only unpicked requests that remain reachable by at least one cooperative agent according to the graph-distance criterion. Define
\begin{equation}
\mathcal{R}^{\mathrm{cf}}_t
=
\left\{
r\in\bar{\mathcal{R}}_t:
\phi_r=0,\;
\exists c\in\mathcal{C}_t
\;\mathrm{s.t.}\;
d_{\mathcal{G}}(\nu_t^c,\rho_r)\leq T_r(t)
\right\}
\label{eq:cooperative_feasible_requests}
\end{equation}

The purpose of this restriction is to focus the attack on requests for which adversarial interference can alter an otherwise plausible cooperative service outcome. Requests outside $\mathcal{R}^{\mathrm{cf}}_t$ are not considered as targets because, under the above reachability criterion, no cooperative agent can reach their pickup location before expiration.

An adversarial agent $a\in\mathcal{A}_t$ can target
$r\in\mathcal{R}^{\mathrm{cf}}_t$ only if reporting the request pickup location satisfies the spoofing constraint:
\begin{equation}
d_{\mathcal{G}}(\nu_t^a,\rho_r)
\leq d_{\max}
\label{eq:adversarial_spoofing_feasibility}
\end{equation}
A feasible adversary-request pair $(a,r)$ therefore represents a request whose pickup location can be reported by adversary $a$ without violating the monitor-aware spoofing constraint.

\subsection{Tiered Target Prioritization}
\label{subsec:tier_construction}
Not all feasible requests provide the same opportunity for disruption. In particular, interfering with a request that is already close to being serviced by a cooperative agent can waste cooperative travel effort and induce additional reassignment. We therefore prioritize requests according to their proximity to cooperative service. For each $r\in\mathcal{R}^{\mathrm{cf}}_t$ define,
\begin{equation}
\varepsilon_r(t)
=
\min_{c\in\mathcal{C}_t}
d_{\mathcal{G}}(\nu_t^c,\rho_r)
\label{eq:cooperative_proximity}
\end{equation}
Smaller values of $\varepsilon_r(t)$ correspond to requests whose pickup locations are closer to at least one cooperative agent. Let $e_0(t)<e_1(t)<\cdots<e_{K_t}(t)$ denote the ordered distinct values of $\varepsilon_r(t)$ over
$r\in\mathcal{R}^{\mathrm{cf}}_t$. We partition the candidate requests into
distance tiers,
\begin{equation}
\mathcal{D}^{(k)}_t
=
\left\{
r\in\mathcal{R}^{\mathrm{cf}}_t:
\varepsilon_r(t)=e_k(t)
\right\},
\qquad
k=0,\ldots,K_t
\label{eq:distance_tiers}
\end{equation}
and define the corresponding tier index,
\begin{equation}
\alpha_r(t)=k
\qquad
\text{if}
\qquad
r\in\mathcal{D}^{(k)}_t
\label{eq:tier_index}
\end{equation}
Lower values of $\alpha_r(t)$ therefore identify targets closer to cooperative service and are assigned higher adversarial priority.

\subsection{Coordinated Adversarial Matching}
\label{subsec:adversarial_matching}
The adversarial team coordinates its targets through a bipartite matching between active adversaries and cooperative-feasible requests. Let $x_{ar}(t)\in\{0,1\}$ indicate whether adversarial agent $a\in\mathcal{A}_t$ targets request
$r\in\mathcal{R}^{\mathrm{cf}}_t$. Each adversary targets at most one request, and each request is targeted by at most one adversary:
\begin{equation}
\sum_{r\in\mathcal{R}^{\mathrm{cf}}_t}x_{ar}(t)\leq 1,
\qquad
\forall a\in\mathcal{A}_t
\label{eq:adversary_one_request_constraint}
\end{equation}
\begin{equation}
\sum_{a\in\mathcal{A}_t}x_{ar}(t)\leq 1,
\qquad
\forall r\in\mathcal{R}^{\mathrm{cf}}_t
\label{eq:request_one_adversary_constraint}
\end{equation}
Only pairs satisfying~\eqref{eq:adversarial_spoofing_feasibility} are included as edges in the matching graph.

The attack follows three lexicographic priorities. It first maximizes the number of requests that can be targeted simultaneously. Among maximum-cardinality matchings, it prioritizes requests in lower distance tiers, and among matchings with identical cardinality and tier priority, it minimizes the total localization displacement required to realize the attack. Equivalently, the adversarial team solves,
\begin{equation}
\begin{aligned}
\operatorname{lexmin}_{x_{ar}(t)}
\Bigg(
&-\sum_{a,r} x_{ar}(t), \\
&\sum_{a,r} \alpha_r(t)x_{ar}(t), \\
&\sum_{a,r}
d_{\mathcal{G}}\!\left(\nu_t^a,\rho_r\right)x_{ar}(t)
\Bigg)
\end{aligned}
\label{eq:adversarial_lexicographic_objective}
\end{equation}
subject to~\eqref{eq:adversary_one_request_constraint}-
\eqref{eq:request_one_adversary_constraint} and the feasible-edge restriction, where the sums are over
$a\in\mathcal{A}_t$ and $r\in\mathcal{R}^{\mathrm{cf}}_t$.

This ordering captures the intended influence-detectability behavior. The attack first seeks to create as many opportunities for assignment manipulation as its spoofing range permits. It then favors requests for which cooperative service is more imminent and, only after these objectives are fixed, selects the matching requiring the smallest aggregate spoofing displacement.

For implementation, the lexicographic problem can be represented as a single weighted assignment problem. Let
\begin{equation}
\bar{d}_t
=
\max_{\substack{
a\in\mathcal{A}_t,\,
r\in\mathcal{R}^{\mathrm{cf}}_t\\
d_{\mathcal{G}}(\nu_t^a,\rho_r)\leq d_{\max}
}}
d_{\mathcal{G}}(\nu_t^a,\rho_r)
\label{eq:max_feasible_adversarial_distance}
\end{equation}
be the largest feasible spoofing distance,
\begin{equation}
\alpha_{\max}(t)
=
\max_{r\in\mathcal{R}^{\mathrm{cf}}_t}\alpha_r(t)
\label{eq:max_tier_index}
\end{equation}
and let $m_t=
\min\left\{
|\mathcal{A}_t|,
|\mathcal{R}^{\mathrm{cf}}_t|
\right\}$
bound the number of matched pairs. Sufficient weights preserving the ordering in~\eqref{eq:adversarial_lexicographic_objective} satisfy, $\beta_t>m_t\bar{d}_t$
and
$\kappa_t>
m_t
\left(
\beta_t\alpha_{\max}(t)+\bar{d}_t
\right)$. 
The weighted matching is then,
\begin{equation}
\min_{x_{ar}(t)}
\sum_{a\in\mathcal{A}_t}
\sum_{r\in\mathcal{R}^{\mathrm{cf}}_t}
\left[
-\kappa_t
+
\beta_t\alpha_r(t)
+
d_{\mathcal{G}}(\nu_t^a,\rho_r)
\right]x_{ar}(t)
\label{eq:single_objective_adversarial_matching}
\end{equation}
subject to the matching and feasibility constraints above.

\subsection{Online Spoofing Procedure}
\label{subsec:adversarial_spoofing_algorithm}
Algorithm~\ref{alg:distance_constrained_spoofing} summarizes the online attack. At each decision epoch, the adversarial team identifies candidate requests, prioritizes them according to cooperative proximity, solves the coordinated matching problem, and generates the localization reports presented to the system. If adversarial agent $a$ is matched to request $r$, it reports the request pickup location $\rho_r$. Otherwise, it reports its true location.

\begin{algorithm}[t]
\caption{Monitor-Aware Distance-Constrained Spoofing}
\label{alg:distance_constrained_spoofing}
\begin{algorithmic}[1]
\REQUIRE Active adversarial fleet $\mathcal{A}_t$,
active cooperative fleet $\mathcal{C}_t$,
outstanding requests $\bar{\mathcal{R}}_t$,
agent locations,
spoofing bound $d_{\max}$
\ENSURE Adversarial reports
$\{\hat{\nu}_t^a\}_{a\in\mathcal{A}_t}$

\STATE Construct $\mathcal{R}^{\mathrm{cf}}_t$ using
\eqref{eq:cooperative_feasible_requests}

\IF{$\mathcal{R}^{\mathrm{cf}}_t=\emptyset$}
    \STATE Set $\hat{\nu}_t^a\gets\nu_t^a$
    for all $a\in\mathcal{A}_t$
    \STATE \textbf{return}
\ENDIF

\STATE Compute $\varepsilon_r(t)$ for all
$r\in\mathcal{R}^{\mathrm{cf}}_t$
using~\eqref{eq:cooperative_proximity}

\STATE Construct tier indices $\alpha_r(t)$ using
\eqref{eq:distance_tiers}-\eqref{eq:tier_index}

\STATE Construct feasible adversary-request edges satisfying
\eqref{eq:adversarial_spoofing_feasibility}

\IF{no feasible edge exists}
    \STATE Set $\hat{\nu}_t^a\gets\nu_t^a$
    for all $a\in\mathcal{A}_t$
    \STATE \textbf{return}
\ENDIF

\STATE Solve the adversarial matching problem
\eqref{eq:single_objective_adversarial_matching}

\FOR{each $a\in\mathcal{A}_t$}
    \IF{$a$ is matched to request $r$}
        \STATE Set $\hat{\nu}_t^a\gets\rho_r$
    \ELSE
        \STATE Set $\hat{\nu}_t^a\gets\nu_t^a$
    \ENDIF
\ENDFOR
\end{algorithmic}
\end{algorithm}

The resulting reports enter the perceived state $\hat{x}_t$ before the trust update and routing decision. The attack is recomputed at every decision epoch, allowing the adversarial team to adapt to the current fleet configuration, outstanding demand, request lifetimes, and spoofing constraints. Reducing $d_{\max}$ simultaneously restricts the set of requests that can be targeted and decreases the localization discrepancy exposed to the monitor, thereby providing the controlled influence-detectability trade-off used in the experimental evaluation.

\section{Trust Evidence and Online Monitoring}
\label{sec:trust_monitor}
The trust-aware framework of Section~\ref{sec:trust_aware_framework} requires observable evidence for determining which agents should remain part of the trusted planning state. In the localization-spoofing setting considered here, we use two complementary evidence sources. \emph{Localization trust} evaluates the reliability of the position information supplied to the planner, whereas \emph{behavioral trust} evaluates whether an agent's participation in task allocation is consistent with cooperative execution. The first source is informative when spoofing produces observable localization anomalies. The second provides complementary evidence when a monitor-aware adversary limits its localization deviation but continues to disrupt service.

At each decision epoch, the centralized monitor uses the currently available localization-integrity observations together with behavioral evidence generated by previously completed or expired requests. Each evidence branch independently produces a local classification in $\{\mathsf{C},\mathsf{S},\mathsf{A}\}$, corresponding to \emph{cooperative}, \emph{suspect}, and \emph{adversarial}. These classifications are fused into the agent state $\chi_i(t)$ introduced in Section~\ref{sec:trust_aware_framework}. Agents for which sufficient adversarial evidence has accumulated are removed before the subsequent routing decision, thereby updating the trusted active fleet and its
corresponding planning state.

\begin{figure*}[t]
    \centering
    \includegraphics[
        width=0.99\linewidth,
        height=0.45\textheight,
        keepaspectratio
    ]{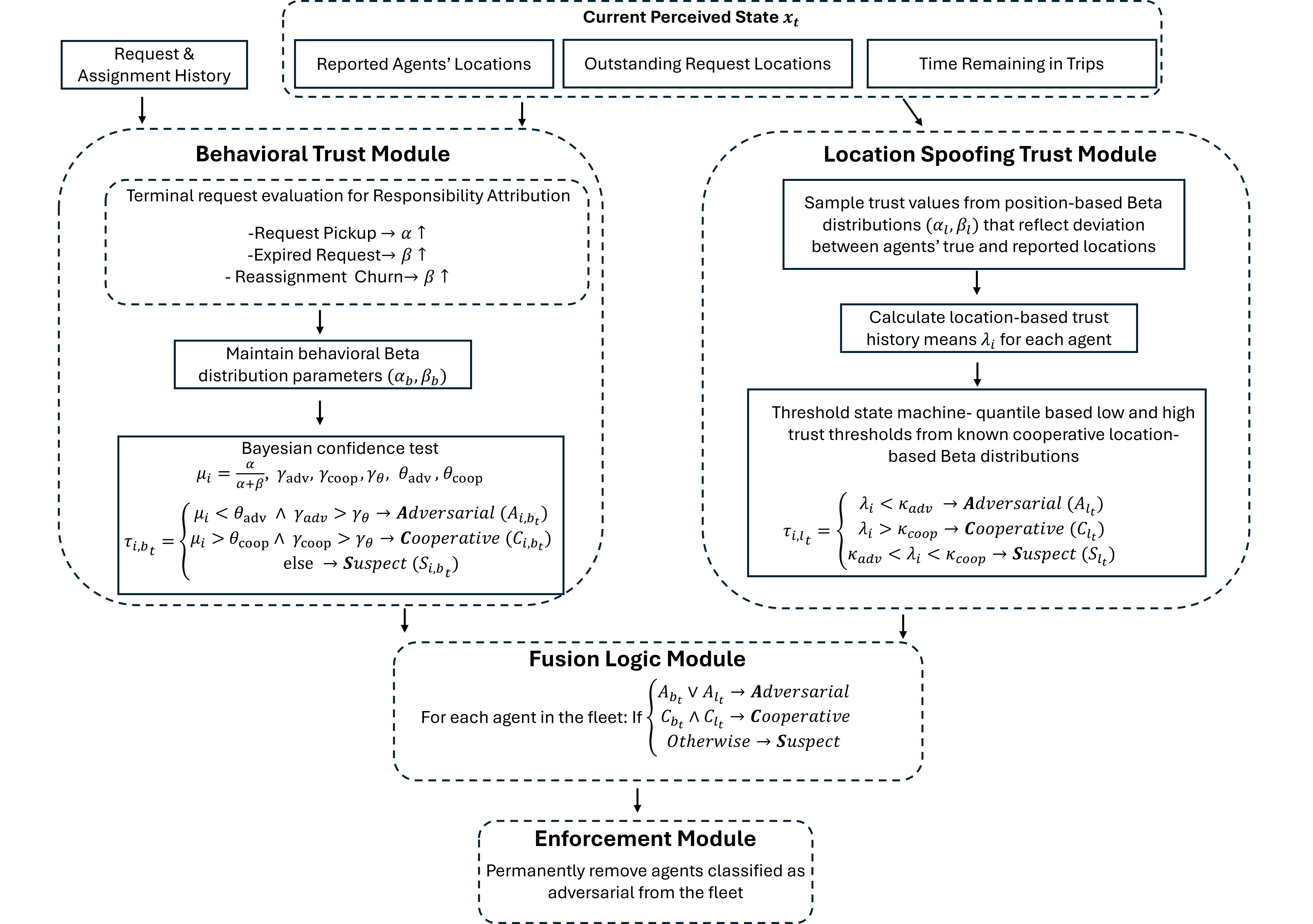}
    \vspace{-5pt}
    \caption{Trust evidence and online monitoring for resilient routing. Localization trust evaluates the reliability of reported position information, while behavioral trust evaluates request-level execution outcomes. The two evidence branches are fused into an agent classification $\chi_i(t)$. Agents classified as adversarial are removed before subsequent planning, producing the trusted active fleet $\mathcal{L}_{t^+}$ and corresponding trusted planning state.}
    \vspace{-7pt}
    \label{fig:trust_based_monitor_block_diagram}
\end{figure*}

\subsection{Localization-Trust Evidence}
\label{subsec:localization_trust}

Localization trust represents confidence in the position information associated with an agent. Let $z^{\mathrm{loc}}_{i,t}\in[0,1]$ denote the localization-trust observation for agent $i$ at time $t$, where larger values indicate greater confidence in the reported localization information. We model localization trust probabilistically using Beta distributions,
\begin{equation}
    Z^{\mathrm{loc}}_{i,t}
    \sim
    \mathrm{Beta}
    \left(
    \alpha^{\mathrm{loc}}_{i,t},
    \beta^{\mathrm{loc}}_{i,t}
    \right)
    \label{eq:localization_trust_distribution}
\end{equation}
whose bounded support and flexible shape provide a convenient representation of both confidence and uncertainty \cite{josang2002beta,teacy2006travos,pippin2012performance}.

The localization-trust model is constructed from real-world GPS spoofing data in two stages. First, localization-related signal features are used to estimate empirical confidence distributions associated with cooperative and spoofed operation. Second, these empirical distributions are related to localization discrepancy so that the trust model varies continuously with the severity of the reported position deviation.

\subsubsection{Calibration From GPS Spoofing Data}
\label{subsubsec:localization_calibration}
Let $s\in\mathbb{R}^d$ denote a normalized feature vector extracted from localization-related signal measurements, and let $y\in\{1,\ldots,K\}$ denote the corresponding operating class. In the datasets used in this work, the classes represent cooperative operation and spoofing strategies with different levels of stealth. We model each class using a class-conditional Gaussian distribution,
\begin{equation}
    p(s\mid y=k)
    =
    \mathcal{N}
    \left(
    s\mid\boldsymbol{\mu}_k,\boldsymbol{\Sigma}_k
    \right)
    \label{eq:class_conditional_localization}
\end{equation}
with class prior
\begin{equation}
    \varpi_k = P(y=k),
    \qquad
    \sum_{k=1}^{K}\varpi_k=1
    \label{eq:localization_class_prior}
\end{equation}
Here, $\varpi_k$ is used for the class prior to distinguish it from the routing-policy notation $\pi_t$. The parameters $\{\varpi_k,\boldsymbol{\mu}_k,\boldsymbol{\Sigma}_k\}_{k=1}^{K}$ are estimated from labeled GPS data. Quadratic discriminant analysis then provides posterior class probabilities,

\begin{equation}
    \widehat{p}_k(s)
    =
    P(y=k\mid s)
    =
    \frac{
    \widehat{\varpi}_k
    \mathcal{N}
    \left(
    s\mid
    \widehat{\boldsymbol{\mu}}_k,
    \widehat{\boldsymbol{\Sigma}}_k
    \right)
    }{
    \displaystyle
    \sum_{j=1}^{K}
    \widehat{\varpi}_j
    \mathcal{N}
    \left(
    s\mid
    \widehat{\boldsymbol{\mu}}_j,
    \widehat{\boldsymbol{\Sigma}}_j
    \right)
    }
    \label{eq:localization_posterior}
\end{equation}

\begin{figure}[t]
    \centering
    \includegraphics[width=0.8\linewidth]
    {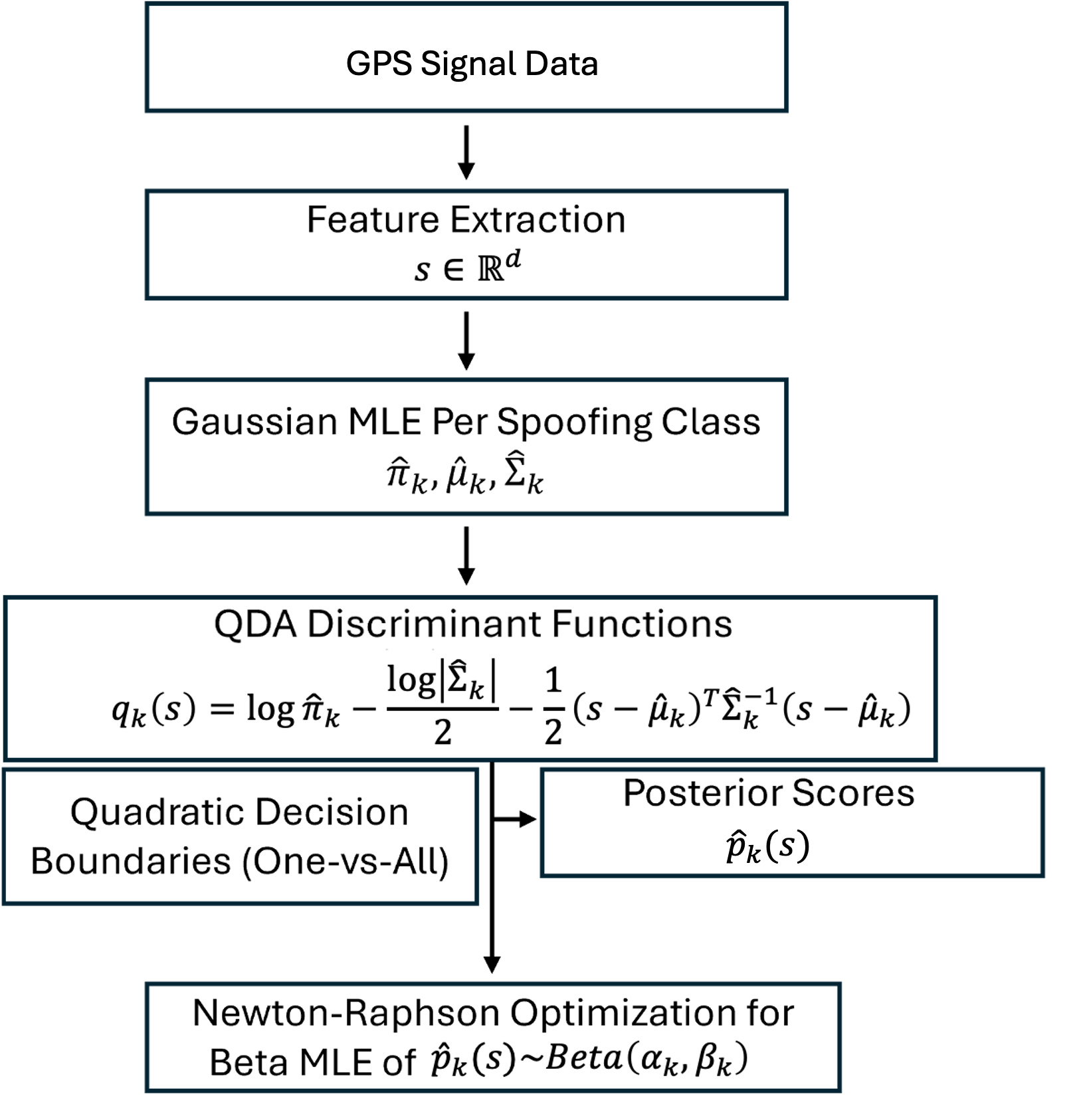}
    \vspace{-7pt}
    \caption{Calibration of localization trust from  real GPS spoofing data. Localization-related features are modeled using class-conditional Gaussian distributions, QDA provides posterior confidence scores, and the resulting scores are represented by class-specific Beta distributions.}
    \label{fig:qda-beta-pipeline}
    \vspace{-7pt}
\end{figure}

The resulting posterior scores lie in $[0,1]$ and are interpreted as localization-confidence observations. For each operating class $k$, we model the corresponding scores using a Beta distribution,
\begin{equation}
    \widehat{p}_k(s)
    \sim
    \mathrm{Beta}(\alpha_k,\beta_k)
    \label{eq:calibrated_beta_distribution}
\end{equation}
The Beta parameters are estimated by maximum likelihood from the posterior scores associated with each class. The complete calibration
pipeline is illustrated in Fig.~\ref{fig:qda-beta-pipeline}. The standard Gaussian and Beta maximum-likelihood estimators are omitted here for brevity.

The resulting distributions characterize the localization-confidence values observed under cooperative operation and different spoofing strategies. In the online model below, we use the cooperative distribution and the distribution associated with the most stealthy spoofing class as endpoints for relating localization discrepancy to trust.

\subsubsection{Discrepancy-Dependent Localization Trust}
\label{subsubsec:localization_model}
The monitor-aware attack of Section~\ref{sec:adversarial_spoofing} explicitly varies the distance between an adversary's reported and actual locations. To capture the corresponding effect on localization trust, define the normalized localization discrepancy as,
\begin{equation}
\Delta_{i,t}
=
\frac{
d_{\mathcal{G}}
\left(
\nu^{i,\mathrm{ref}}_t,
\hat{\nu}^i_t
\right)
}{
D(\mathcal{G})
},
\qquad
\Delta_{i,t}\in[0,1]
\label{eq:normalized_spoofing_discrepancy_monitor}
\end{equation}
where $\hat{\nu}^i_t$ is the reported location and $\nu^{i,\mathrm{ref}}_t$ is a trusted or independently validated localization reference. In simulation, $\nu^{i,\mathrm{ref}}_t=\nu^i_t$. A deployed monitor does not need to have direct access to the true physical
position as the reference may instead be supplied by an independent localization-integrity mechanism, redundant sensing, or another trusted source.

Let $(\mu_l,v_l)$ denote the mean and variance of the calibrated cooperative localization-trust distribution and $(\mu_a,v_a)$ denote the corresponding quantities for the selected adversarial distribution. We interpolate these moments according to the observed localization discrepancy:
\begin{equation}
    \mu^{\mathrm{loc}}_{i,t}
    =
    (\mu_l-\mu_a)(1-\Delta_{i,t})+\mu_a
    \label{eq:spoofing_adjusted_mean_monitor}
    \end{equation}
    \begin{equation}
    v^{\mathrm{loc}}_{i,t}
    =
    (v_l-v_a)(1-\Delta_{i,t})+v_a
    \label{eq:spoofing_adjusted_variance_monitor}
\end{equation}
Thus, when $\Delta_{i,t}=0$, the model coincides with the cooperative trust distribution. Increasing the discrepancy progressively shifts the distribution toward the calibrated adversarial model. The corresponding Beta parameters are obtained by moment matching:
\begin{equation}
    \alpha^{\mathrm{loc}}_{i,t}
    =
    \mu^{\mathrm{loc}}_{i,t}
    \left[
    \frac{
    \mu^{\mathrm{loc}}_{i,t}
    \left(1-\mu^{\mathrm{loc}}_{i,t}\right)
    }{
    v^{\mathrm{loc}}_{i,t}
    }
    -1
    \right]
    \label{eq:beta_alpha_from_moments_monitor}
\end{equation}
\begin{equation}
    \beta^{\mathrm{loc}}_{i,t}
    =
    \left(1-\mu^{\mathrm{loc}}_{i,t}\right)
    \left[
    \frac{
    \mu^{\mathrm{loc}}_{i,t}
    \left(1-\mu^{\mathrm{loc}}_{i,t}\right)
    }{
    v^{\mathrm{loc}}_{i,t}
    }
    -1
    \right]
    \label{eq:beta_beta_from_moments_monitor}
\end{equation}
These expressions are valid when,
\begin{equation}
    0<\mu^{\mathrm{loc}}_{i,t}<1,
    \qquad
    0<v^{\mathrm{loc}}_{i,t}
    <
    \mu^{\mathrm{loc}}_{i,t}
    \left(1-\mu^{\mathrm{loc}}_{i,t}\right)
\end{equation}
This construction captures the influence-detectability trade-off introduced in Section~\ref{sec:adversarial_spoofing}: smaller spoofing deviations produce localization-trust observations closer to nominal operation, whereas larger deviations provide stronger localization-based evidence of manipulation.
In simulation, the online localization-trust observation is sampled as,
\begin{equation}
    z^{\mathrm{loc}}_{i,t}
    \sim
    \mathrm{Beta}
    \left(
    \alpha^{\mathrm{loc}}_{i,t},
    \beta^{\mathrm{loc}}_{i,t}
    \right)
    \label{eq:location_trust_sample_monitor}
\end{equation}
In deployment, $z^{\mathrm{loc}}_{i,t}$ may instead be supplied directly by a localization-integrity estimator. The online monitor therefore operates on bounded localization-trust observations and does not require direct access to the true state used by the simulator.

\subsubsection{Temporal Aggregation and Localization Classification}
\label{subsubsec:localization_classification}
To reduce sensitivity to individual observations, the monitor aggregates localization trust over a sliding window of length $W_{\mathrm{loc}}$. Define
\begin{equation}
    \mathcal{W}^{\mathrm{loc}}_{i,t}
    =
    \left\{
    k :
    \max\{0,t-W_{\mathrm{loc}}+1\}\leq k\leq t,
    \;
    i\in\mathcal{L}_k
    \right\}
    \label{eq:location_trust_window}
\end{equation}
and let $n^{\mathrm{loc}}_{i,t}
    =
    \left|
    \mathcal{W}^{\mathrm{loc}}_{i,t}
    \right|$. The aggregated localization-trust score is,
\begin{equation}
    \bar{z}^{\mathrm{loc}}_{i,t}
    =
    \frac{1}{n^{\mathrm{loc}}_{i,t}}
    \sum_{k\in\mathcal{W}^{\mathrm{loc}}_{i,t}}
    z^{\mathrm{loc}}_{i,k}
    \label{eq:location_trust_window_average}
\end{equation}
Given thresholds
$\kappa^{\mathrm{loc}}_{\mathrm{adv}}$
and
$\kappa^{\mathrm{loc}}_{\mathrm{coop}}$, satisfying $\kappa^{\mathrm{loc}}_{\mathrm{adv}}
<
\kappa^{\mathrm{loc}}_{\mathrm{coop}}$ the localization branch produces,
\begin{equation}
\chi^{\mathrm{loc}}_i(t)
=
\begin{cases}
\mathsf{S},
&
n^{\mathrm{loc}}_{i,t}<W_{\mathrm{loc}},
\\[1mm]
\mathsf{A},
&
\bar{z}^{\mathrm{loc}}_{i,t}
<
\kappa^{\mathrm{loc}}_{\mathrm{adv}},
\\[1mm]
\mathsf{C},
&
\bar{z}^{\mathrm{loc}}_{i,t}
>
\kappa^{\mathrm{loc}}_{\mathrm{coop}},
\\[1mm]
\mathsf{S},
&
\text{otherwise}
\end{cases}
\label{eq:location_module_classification}
\end{equation}

All agents remain in the suspect state until a complete localization window has been accumulated. The adversarial threshold is selected conservatively from nominal cooperative data, while the cooperative threshold characterizes typical nominal localization behavior.

Localization evidence is most informative when spoofing creates sufficiently large observable discrepancies. Under the monitor-aware strategy, however, an adversary may reduce its spoofing magnitude and remain statistically close to cooperative operation. This motivates the second, execution-based evidence branch.

\subsection{Behavioral Trust and Request Responsibility}
\label{subsec:behavioral_trust}
Behavioral trust evaluates whether an agent's participation in task allocation is consistent with cooperative service. Unlike localization trust, it does not directly evaluate the reported position. Instead, it uses request-level outcomes that become observable as the routing process evolves. This allows the monitor to accumulate evidence against agents that generate relatively benign localization observations but repeatedly fail to produce the service outcomes anticipated by the planner.

For each request $r$, the monitor maintains an assignment history
\begin{equation}
    \mathcal{H}_r
    =
    \left\{
    \left(
    i_m,
    t^m_{\mathrm{assign}},
    \hat{t}^m_{\mathrm{pickup}},
    t^m_{\mathrm{unassign}}
    \right)
    \right\}_{m=1}^{M_r}
    \label{eq:request_assignment_history}
\end{equation}
where $i_m$ is the agent assigned during interval $m$, $t^m_{\mathrm{assign}}$ is the assignment time, $\hat{t}^m_{\mathrm{pickup}}$ is the predicted pickup time associated with that assignment, and $t^m_{\mathrm{unassign}}$ records when the assignment ends if the request is reassigned. A request reaches a \emph{terminal event} at time $t^r_{\mathrm{term}}$ through one of two outcomes:
\begin{enumerate}
    \item \emph{Successful pickup:}
    $t^r_{\mathrm{term}}=t^r_{\mathrm{pickup}}$
    \item \emph{Request expiration:}
    $t^r_{\mathrm{term}}=t^r_{\mathrm{exp}}$
\end{enumerate}
Behavioral evidence is updated only after a terminal event, allowing the monitor to reconstruct the complete assignment history before assigning responsibility. Each agent maintains a behavioral trust distribution,

\begin{figure}[t]
    \centering
    \includegraphics[
        width=0.99\linewidth,
        height=0.4\textheight,
        keepaspectratio
    ]{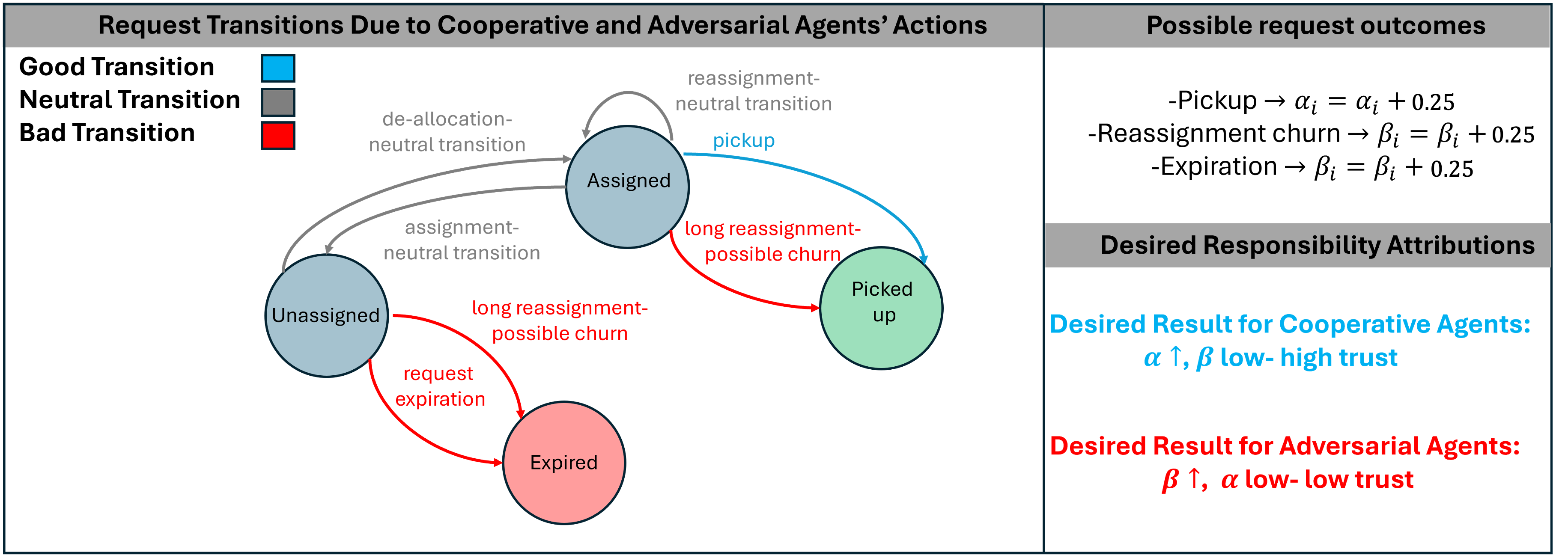}
    \vspace{-5pt}
    \caption{Request-level responsibility attribution for behavioral trust. After a request reaches a terminal event, its assignment history is used to attribute successful service, expiration, and harmful reassignment churn to the agents involved.}
    \vspace{-7pt}
    \label{fig:request_attribution_module}
\end{figure}

\begin{equation}
    Z^{\mathrm{beh}}_i(t)
    \sim
    \mathrm{Beta}
    \left(
    \alpha^{\mathrm{beh}}_i(t),
    \beta^{\mathrm{beh}}_i(t)
    \right)
    \label{eq:behavioral_trust_distribution}
\end{equation}
where $\alpha^{\mathrm{beh}}_i(t)$ accumulates evidence of compliant behavior and $\beta^{\mathrm{beh}}_i(t)$ accumulates evidence of harmful behavior. The parameters are initialized as
\begin{equation}
    \alpha^{\mathrm{beh}}_i(0)=\alpha^{\mathrm{beh}}_0,
    \qquad
    \beta^{\mathrm{beh}}_i(0)=\beta^{\mathrm{beh}}_0
\end{equation}
If request $r$ is successfully picked up by agent $i_{\mathrm{pick}}$, the servicing agent receives positive evidence:
\begin{equation}
    \alpha^{\mathrm{beh}}_{i_{\mathrm{pick}}}
    \leftarrow
    \alpha^{\mathrm{beh}}_{i_{\mathrm{pick}}}
    +
    w_{\mathrm{pick}}
    \label{eq:behavior_success_update}
\end{equation}
If a request expires, its most recently assigned agent receives an expiration penalty:
\begin{equation}
    \beta^{\mathrm{beh}}_{i_{\mathrm{last}}}
    \leftarrow
    \beta^{\mathrm{beh}}_{i_{\mathrm{last}}}
    +
    w_{\mathrm{exp}}
    \label{eq:behavior_expiration_update}
\end{equation}
We additionally attribute harmful reassignment churn using the complete request history. Define,

\begin{align}
    \mathcal{Q}_r
    =
    \bigl\{
    i_m :\;&
    \left(
    i_m,
    t^m_{\mathrm{assign}},
    \hat{t}^m_{\mathrm{pickup}},
    t^m_{\mathrm{unassign}}
    \right)
    \in\mathcal{H}_r,
    \nonumber\\
    &
    \hat{t}^m_{\mathrm{pickup}}
    <
    t^r_{\mathrm{term}},
    \;
    i_m\neq i_{\mathrm{pick}}
    \bigr\}
    \label{eq:churn_responsible_set}
\end{align}
where the condition $i_m\neq i_{\mathrm{pick}}$ is omitted when the request expires. Each agent in $\mathcal{Q}_r$ receives the penalty,
\begin{equation}
    \beta^{\mathrm{beh}}_i
    \leftarrow
    \beta^{\mathrm{beh}}_i
    +
    w_{\mathrm{churn}},
    \qquad
    \forall i\in\mathcal{Q}_r
    \label{eq:behavior_churn_update}
\end{equation}
This rule captures agents whose anticipated pickup should have occurred before the request terminated but who did not complete the service. Using the complete assignment history is important in a reassignment-based system because an agent may repeatedly disrupt requests without being the final assignee at expiration. The posterior mean behavioral trust is the corresponding Beta distribution's mean,
\begin{equation}
    \bar{z}^{\mathrm{beh}}_i(t)
    =
    \frac{
    \alpha^{\mathrm{beh}}_i(t)
    }{
    \alpha^{\mathrm{beh}}_i(t)
    +
    \beta^{\mathrm{beh}}_i(t)
    }
    \label{eq:behavioral_posterior_mean}
\end{equation}
To avoid classification from sparse behavioral evidence, define the accumulated behavioral evidence weight as,
\begin{equation}
    n^{\mathrm{beh}}_i(t)
    =
    \alpha^{\mathrm{beh}}_i(t)
    +
    \beta^{\mathrm{beh}}_i(t)
    -
    \alpha^{\mathrm{beh}}_0
    -
    \beta^{\mathrm{beh}}_0
    \label{eq:behavioral_evidence_weight}
\end{equation}
We also define the posterior-confidence quantities,
\begin{equation}
    P^{\mathrm{adv}}_i(t)
    =
    \Pr\left[
    Z^{\mathrm{beh}}_i(t)
    \leq
    \theta^{\mathrm{beh}}_{\mathrm{adv}}
    \right]
    \label{eq:behavioral_adversarial_probability}
\end{equation}

\begin{equation}
    P^{\mathrm{coop}}_i(t)
    =
    \Pr\left[
    Z^{\mathrm{beh}}_i(t)
    \geq
    \theta^{\mathrm{beh}}_{\mathrm{coop}}
    \right]
    \label{eq:behavioral_cooperative_probability}
\end{equation}
The behavioral branch produces the classification,
\begin{equation}
    \chi^{\mathrm{beh}}_i(t)
    =
    \begin{cases}
    \mathsf{S},
    &
    n^{\mathrm{beh}}_i(t)
    <
    n^{\mathrm{beh}}_{\min},
    \\[1mm]
    \mathsf{A},
    &
    \bar{z}^{\mathrm{beh}}_i(t)
    \leq
    \theta^{\mathrm{beh}}_{\mathrm{adv}}
    \;\land\;
    P^{\mathrm{adv}}_i(t)
    \geq
    \gamma^{\mathrm{beh}}_{\mathrm{adv}},
    \\[1mm]
    \mathsf{C},
    &
    \bar{z}^{\mathrm{beh}}_i(t)
    \geq
    \theta^{\mathrm{beh}}_{\mathrm{coop}}
    \;\land\;
    P^{\mathrm{coop}}_i(t)
    \geq
    \gamma^{\mathrm{beh}}_{\mathrm{coop}},
    \\[1mm]
    \mathsf{S},
    &
    \text{otherwise}
    \end{cases}
    \label{eq:behavior_module_classification}
\end{equation}
Thus, behavioral classification requires both sufficient accumulated evidence and sufficient posterior confidence; ambiguous agents remain suspect. Algorithm~\ref{alg:behavioral_trust_update} summarizes the behavioral update procedure.

\begin{algorithm}[t]
\caption{Behavioral Trust Update}
\label{alg:behavioral_trust_update}
\begin{algorithmic}[1]
\REQUIRE Newly terminal requests $\mathcal{T}_t$,
request histories $\{\mathcal{H}_r\}$,
behavioral trust parameters
\FOR{each $r\in\mathcal{T}_t$}
    \STATE Recover the complete assignment history $\mathcal{H}_r$
    \IF{$r$ was successfully picked up}
        \STATE Reward the servicing agent using
        \eqref{eq:behavior_success_update}
    \ELSIF{$r$ expired}
        \STATE Penalize the last assigned agent using
        \eqref{eq:behavior_expiration_update}
    \ENDIF
    \STATE Construct $\mathcal{Q}_r$ using
    \eqref{eq:churn_responsible_set}
    \STATE Penalize agents in $\mathcal{Q}_r$ using
    \eqref{eq:behavior_churn_update}
    \STATE Mark $r$ as processed
\ENDFOR
\end{algorithmic}
\end{algorithm}

\subsection{Trust Fusion and Agent Classification}
\label{subsec:fusion_logic}
Localization and behavioral trust provide evidence about different parts of the planner-execution relationship. Localization trust evaluates whether the information entering the planning state is consistent with an independent localization reference. Behavioral trust evaluates whether task-level execution is consistent with the service behavior expected of a cooperative agent. Their complementarity is particularly important for monitor-aware attacks: reducing localization discrepancy may weaken the first source of evidence but does not necessarily eliminate the downstream behavioral consequences of repeatedly attracting and failing to service requests.

The local classifications are fused according to,

\begin{equation}
    \chi_i(t)
    =
    \begin{cases}
    \mathsf{A},
    &
    \chi^{\mathrm{loc}}_i(t)=\mathsf{A}
    \;\lor\;
    \chi^{\mathrm{beh}}_i(t)=\mathsf{A},
    \\[1mm]
    \mathsf{C},
    &
    \chi^{\mathrm{loc}}_i(t)=\mathsf{C}
    \;\land\;
    \chi^{\mathrm{beh}}_i(t)=\mathsf{C},
    \\[1mm]
    \mathsf{S},
    &
    \text{otherwise}
    \end{cases}
    \label{eq:fusion_rule}
\end{equation}

Thus, sufficiently strong evidence from either branch is sufficient for adversarial classification, whereas cooperative classification requires agreement between both branches. Conflicting or incomplete evidence leaves the agent in the suspect state.

The suspect state is important because classification has a direct planning consequence. Suspect agents remain under consideration during planning while additional evidence is collected. This avoids forcing a binary decision from insufficient evidence, although it also means that planner-execution mismatch may persist during the detection interval.

\subsection{Enforcement and Trusted-Fleet Update}
\label{subsec:enforcement}
The fused classification becomes operational through the trusted-state construction introduced in Section~\ref{subsec:trusted_planning_state}. If $\chi_i(t)=\mathsf{A}$, agent $i$ is excluded from subsequent planning. The active planning fleet is therefore updated according to,
\begin{equation}
    \mathcal{L}_{t^+}
    =
    \mathcal{L}_t
    \setminus
    \left\{
    i\in\mathcal{L}_t :
    \chi_i(t)=\mathsf{A}
    \right\}
    \label{eq:active_fleet_removal}
\end{equation}
where $t^+$ denotes the instant after monitoring and enforcement at decision epoch $t$. In the implementation considered, removal is permanent:
\begin{equation}
    \chi_i(t)=\mathsf{A}
    \quad\Longrightarrow\quad
    i\notin\mathcal{L}_{t'},
    \qquad
    \forall t'>t
    \label{eq:permanent_removal}
\end{equation}
Any outstanding assignment involving a removed agent is revoked, and the corresponding request history is retained so that subsequent behavioral attribution remains consistent. The perceived state is then restricted to the remaining fleet,
$\hat{x}_{t^+}
=
\left.
\hat{x}_t
\right|_{\mathcal{L}_{t^+}}$ which is the routing-specific realization of the trusted planning state introduced in Section~\ref{sec:trust_aware_framework}.


This enforcement step provides the connection between trust estimation and sequential planning. The monitor does not modify the routing objective itself. Instead, it modifies the set of agents considered by the planner. While undetected adversaries remain in $\mathcal{L}_{t^+}$, planner-execution mismatch can persist. As sufficient localization or behavioral evidence accumulates and those agents are removed, the agents considered during planning become better aligned with those expected to execute cooperative routing commands. Section~\ref{sec:trust_aware_rollout} examines the consequences of this alignment for rollout-based planning.

\section{Trust-Aware Rollout Planning}
\label{sec:trust_aware_rollout}
The preceding sections determine which agents remain in the trusted planning state but do not prescribe how routing decisions should be computed from that state. We now instantiate the planning component of the framework using rollout. Rollout is particularly useful for exposing the effect of planner-execution mismatch because its decision quality depends on simulated future trajectories accurately representing the system that will execute the selected actions.

We use Instantaneous Assignment with Reassignment (IA-RA) as the base routing policy. The resulting planner first applies the trust monitor to construct the filtered active fleet $\mathcal{L}_{t^+}$ and perceived planning state $\hat{x}_{t^+}$, and then performs rollout over that state. Monitoring does not modify the rollout optimization itself. Instead, it modifies the set of agents represented in the rollout model. While undetected adversaries remain in $\mathcal{L}_{t^+}$, simulated and executed behavior may still differ. As adversarial agents are detected and removed, the set of agents considered during planning becomes better aligned with the agents executing the selected routing commands.

\subsection{IA-RA Base Policy}
\label{subsec:iara_base_policy}
Let $\mathcal{E}_{t^+}\subseteq\mathcal{L}_{t^+}$ denote the set of agents eligible for assignment after monitoring at decision epoch $t$. IA-RA, denoted by $\pi^{\mathrm{IA\text{-}RA}}$, computes assignments between eligible agents and outstanding unpicked requests using the perceived planning state $\hat{x}_{t^+}$. For an eligible agent $i\in\mathcal{E}_{t^+}$ and an outstanding request $r\in\bar{\mathcal{R}}_t$, define the assignment cost as,
\begin{equation}
C_t(i,r)
=
d_{\mathcal{G}}
\left(
\hat{\nu}_{t^+}^{i},
\rho_r
\right)
+
d_{\mathcal{G}}
\left(
\rho_r,
\delta_r
\right)
\label{eq:iara_assignment_cost}
\end{equation}
IA-RA solves the corresponding minimum-cost bipartite matching problem and recomputes assignments as the system evolves. Requests that have not yet been picked up may therefore be reassigned when new demand arrives or when the perceived fleet state changes.

The policy is computationally tractable and provides a natural base policy for rollout-based improvement \cite{garces2024approximate}. In the remainder of this section, IA-RA is used both to initialize the current joint action and to generate future actions within the rollout simulations.

\subsection{Rollout Over the Trusted Planning State}
\label{subsec:rollout_over_trusted_state}
At decision epoch $t$, let $\mathcal{L}_{t^+}$ denote the active fleet after monitoring and enforcement, and let $\hat{x}_{t^+}$ denote the corresponding trusted planning state defined in Section~\ref{subsec:enforcement}. The rollout planner evaluates alternative current routing actions while using IA-RA as the base policy for subsequent simulated decisions. Let
\begin{equation}
\omega_t^m
=
\left\{
\eta_{t+h}^m,
\boldsymbol{\rho}_{t+h}^m,
\boldsymbol{\delta}_{t+h}^m
\right\}_{h=0}^{H-1},
\qquad
m=1,\ldots,M
\label{eq:rollout_scenario}
\end{equation}
denote the $m$-th sampled demand realization over a lookahead horizon of length $H$, where $M$ is the number of Monte Carlo scenarios. For a candidate current action
$u_t\in\mathcal{U}(\hat{x}_{t^+})$,
the rollout simulation uses $u_t^m = u_t$ at the first simulated stage and applies the IA-RA base policy at subsequent stages:
\begin{equation}
u_{t+h}^m
=
\pi^{\mathrm{IA\text{-}RA}}
\left(
\hat{x}_{t+h}^m
\right),
\qquad
h=1,\ldots,H-1
\label{eq:rollout_base_policy}
\end{equation}
The simulated perceived state evolves according to,
\begin{equation}
\hat{x}_{t+h+1}^m
=
\hat{f}
\left(
\hat{x}_{t+h}^m,
u_{t+h}^m,
\eta_{t+h}^m,
\boldsymbol{\rho}_{t+h}^m,
\boldsymbol{\delta}_{t+h}^m
\right)
\label{eq:rollout_simulated_transition}
\end{equation}
where $\hat{f}$ denotes the transition model used by the planner. Within this model, every agent retained in $\mathcal{L}_{t^+}$ is assumed to execute its commanded routing action.

Using the stage-cost function introduced in Section~\ref{sec:problem_formulation}, the finite-horizon cost under scenario $m$ is,
\begin{align}
J_H^m
\left(
\hat{x}_{t^+},
u_t
\right)
={}&
\sum_{h=0}^{H-1}
g_{t+h}
\left(
\hat{x}_{t+h}^m,
u_{t+h}^m,
\eta_{t+h}^m,
\boldsymbol{\rho}_{t+h}^m,
\boldsymbol{\delta}_{t+h}^m
\right)
\nonumber\\
&+
g_T
\left(
\hat{x}_{t+H}^m
\right)
\label{eq:finite_horizon_rollout_cost}
\end{align}
The terminal cost penalizes residual service burden beyond the finite lookahead horizon. In the implementation considered here, we use,
\begin{equation}
g_T
\left(
\hat{x}_{t+H}^m
\right)
=
\left|
\bar{\mathcal{R}}_{t+H}^m
\right|
\label{eq:rollout_terminal_cost}
\end{equation}
The sampled cost-to-go estimate is,
\begin{equation}
\widehat{Q}_H
\left(
\hat{x}_{t^+},
u_t
\right)
=
\frac{1}{M}
\sum_{m=1}^{M}
J_H^m
\left(
\hat{x}_{t^+},
u_t
\right)
\label{eq:rollout_q_estimate}
\end{equation}
A full joint rollout would then select,
\begin{equation}
u_t^{\mathrm{roll}}
=
\underset{
u_t\in\mathcal{U}(\hat{x}_{t^+})
}{\arg\min}
\;
\widehat{Q}_H
\left(
\hat{x}_{t^+},
u_t
\right)
\label{eq:rollout_action}
\end{equation}

For each Monte Carlo scenario, the future demand realization is sampled once and held fixed when comparing candidate current actions. This common-scenario evaluation reduces variation caused solely by different demand samples during action comparison.

\subsection{Planner-Execution Mismatch Under Undetected Adversaries}
\label{subsec:rollout_failure_adversaries}
The rollout model in~\eqref{eq:rollout_simulated_transition} assumes that every agent represented in the trusted planning state follows the action selected by the planner. This assumption is reasonable when the retained agents execute their assigned actions cooperatively. It can fail, however, while an adversarial agent remains undetected. Let $\mathcal{A}_{t^+}
= \mathcal{A}_t \cap \mathcal{L}_{t^+}$ denote the adversarial agents that remain in the set considered for planning after monitoring. 
If $\mathcal{A}_{t^+}\neq\emptyset$, an adversarial agent may be represented during rollout as a service resource that follows the commanded action even though its physical execution differs from this assumption. The simulated transition is,
\begin{equation}
\hat{x}_{k+1}^m
=
\hat{f}
\left(
\hat{x}_k^m,
u_k^m,
\eta_k^m,
\boldsymbol{\rho}_k^m,
\boldsymbol{\delta}_k^m
\right)
\label{eq:rollout_perceived_transition}
\end{equation}
whereas the physical system evolves according to,
\begin{equation}
x_{k+1}
=
f
\left(
x_k,
u_k^{\mathrm{exec}},
\eta_k,
\boldsymbol{\rho}_k,
\boldsymbol{\delta}_k
\right)
\label{eq:rollout_true_transition}
\end{equation}
For an undetected adversarial agent $a\in\mathcal{A}_{t^+}$, it is possible that,
\begin{equation}
u_k^{a,\mathrm{exec}}
\neq
u_k^a
\label{eq:adversarial_execution_mismatch}
\end{equation}

This is the rollout-specific manifestation of the planner-execution mismatch introduced in Section~\ref{sec:problem_formulation}. The planner evaluates candidate actions using trajectories in which all retained agents provide the service assumed by the routing model, whereas the executing system may contain agents that ignore those assignments. Consequently, the estimated cost-to-go $\widehat{Q}_H$ may not represent the cost generated by physical execution, and the action favored by rollout need not improve upon the IA-RA base action in the adversarial system.

This distinction is important when interpreting classical rollout improvement results. Standard rollout arguments compare a rollout policy with its base policy under a common system model \cite{bertsekas2021rollout,bertsekas2021multiagent,bertsekas2023coursenew}. When the model used to evaluate candidate actions differs from the system that executes those actions, the premise supporting that comparison is no longer satisfied. In addition, the finite-horizon Monte Carlo rollout used here is itself an approximation to exact rollout. We therefore evaluate improvement empirically rather than claim a general cost-improvement guarantee under adversarial operation.

\subsection{Trust-Aware One-at-a-Time Rollout}
\label{subsec:trust_aware_one_at_a_time_rollout}
The full optimization in~\eqref{eq:rollout_action} can be computationally expensive because the joint action space grows combinatorially with the number of active agents. We therefore use a one-at-a-time rollout procedure \cite{bertsekas2021rollout}, in which agent actions are improved sequentially while the other components of the current joint action are held fixed.

The procedure is initialized with the IA-RA action for the monitored planning state:
\begin{equation}
u_t^{[0]}
=
\pi^{\mathrm{IA\text{-}RA}}
\left(
\hat{x}_{t^+}
\right)
\label{eq:rollout_initial_action}
\end{equation}
Let the agents in $\mathcal{L}_{t^+}$ be indexed according to the order in which they are processed by the one-at-a-time rollout procedure. Let $N_{t^+}=\left|\mathcal{L}_{t^+}\right|$ denote the number of agents retained after monitoring. For agent $i$, let
$\mathcal{U}_i(\hat{x}_{t^+})$
denote its admissible action set. Suppose that the actions of agents $1,\ldots,i-1$ have already been updated. For a candidate action,
$a_i\in\mathcal{U}_i(\hat{x}_{t^+})$,
define the candidate joint action,
\begin{equation}
u_t^{[i]}(a_i)
=
\left(
u_t^{[i-1],1},
\ldots,
u_t^{[i-1],i-1},
a_i,
u_t^{[i-1],i+1},
\ldots,
u_t^{[i-1],N_{t^+}}
\right)
\label{eq:one_at_a_time_candidate}
\end{equation}
The action selected for agent $i$ is,
\begin{equation}
a_i^*
=
\underset{
a_i\in\mathcal{U}_i(\hat{x}_{t^+})
}{\arg\min}
\;
\widehat{Q}_H
\left(
\hat{x}_{t^+},
u_t^{[i]}(a_i)
\right)
\label{eq:one_at_a_time_action}
\end{equation}
The joint action is then updated according to,
\begin{equation}
u_t^{[i]}
=
u_t^{[i]}(a_i^*)
\label{eq:one_at_a_time_update}
\end{equation}
After all retained agents have been processed, the trust-aware rollout command is,
\begin{equation}
u_t^{\mathrm{TA\text{-}roll}}
=
u_t^{[N_{t^+}]}
\label{eq:trust_aware_rollout_action}
\end{equation}

The adjective \emph{trust-aware} refers to the state on which rollout operates rather than to a modification of the rollout objective. The optimization in~\eqref{eq:one_at_a_time_action} is performed only over the set of agents retained after monitoring. Trust therefore affects rollout by changing which agents and reported states are represented during lookahead.

\subsection{Closed-Loop Trust-Aware Rollout Procedure}
\label{subsec:closed_loop_trust_aware_rollout}
Algorithm~\ref{alg:trust_aware_rollout} summarizes the interaction between monitoring and rollout planning. The ordering follows the decision-epoch sequence introduced in Section~\ref{subsec:closed_loop_decision_epoch}: current reports and available evidence are processed first, the trusted planning state is constructed next, and rollout then computes the routing command. Physical execution subsequently generates evidence for future monitoring updates.

\begin{algorithm}[t]
\caption{Trust-Aware One-at-a-Time Rollout Planning}
\label{alg:trust_aware_rollout}
\begin{algorithmic}[1]
\REQUIRE Perceived state $\hat{x}_t$, active fleet $\mathcal{L}_t$,
monitor state, base policy $\pi^{\mathrm{IA\text{-}RA}}$,
demand model, rollout horizon $H$, number of scenarios $M$
\ENSURE Rollout command $u_t^{\mathrm{TA\text{-}roll}}$
and filtered fleet $\mathcal{L}_{t^+}$

\STATE Process currently available localization and behavioral evidence
\STATE Update $\chi_i(t)$ for all $i\in\mathcal{L}_t$
\STATE Remove all agents satisfying $\chi_i(t)=\mathsf{A}$
\STATE Construct $\mathcal{L}_{t^+}$ and $\hat{x}_{t^+}$
\STATE Set $N_{t^+}\gets|\mathcal{L}_{t^+}|$
\STATE Generate common demand scenarios
$\{\omega_t^m\}_{m=1}^{M}$ over horizon $H$
\STATE Initialize
$u_t^{[0]}\gets
\pi^{\mathrm{IA\text{-}RA}}(\hat{x}_{t^+})$

\FOR{$i=1,\ldots,N_{t^+}$}
    \STATE Construct $\mathcal{U}_i(\hat{x}_{t^+})$
    \FOR{each $a_i\in\mathcal{U}_i(\hat{x}_{t^+})$}
        \STATE Form the candidate joint action $u_t^{[i]}(a_i)$
        \STATE Estimate
        $\widehat{Q}_H
        \left(
        \hat{x}_{t^+},
        u_t^{[i]}(a_i)
        \right)$
    \ENDFOR
    \STATE Select $a_i^*$ according to
    \eqref{eq:one_at_a_time_action}
    \STATE Update $u_t^{[i]}$ according to
    \eqref{eq:one_at_a_time_update}
\ENDFOR

\STATE Set
$u_t^{\mathrm{TA\text{-}roll}}
\gets
u_t^{[N_{t^+}]}$
\STATE \textbf{return}
$u_t^{\mathrm{TA\text{-}roll}}$ and $\mathcal{L}_{t^+}$
\end{algorithmic}
\end{algorithm}

The commanded action is then passed to the physical fleet. Cooperative agents execute their assigned service actions, whereas any adversarial agents that remain undetected may deviate from the commanded action according to the adversarial model of Section~\ref{sec:adversarial_spoofing}. The resulting true-state evolution, request outcomes, and assignment histories generate evidence used at subsequent monitoring epochs.

\subsection{Effect of Trust on Planner-Execution Consistency}
\label{subsec:trust_rollout_consistency}
Trust-aware rollout has two operating regimes. While adversarial agents remain undetected, the filtered fleet may satisfy $\mathcal{A}_{t^+}\neq\emptyset$. 

In this regime, rollout remains vulnerable to planner-execution mismatch because its lookahead model treats the retained adversarial agents as cooperative service resources.

If the monitor subsequently removes all remaining adversarial agents, then $\mathcal{A}_{t^+} = \emptyset$. Up to cooperative-agent false positives, the retained set of agents considered for planning then contains only agents that execute commands according to the cooperative routing model. The specific source of mismatch caused by adversarial non-execution is therefore removed, and the rollout simulations become better aligned with the physical fleet executing the selected actions.

This does not imply that monitoring eliminates all modeling errors or that finite-horizon rollout is guaranteed to outperform its base policy at every decision epoch. Demand sampling, finite lookahead, model approximation, and cooperative-agent false positives may still affect performance. Rather, adversarial removal addresses the specific planner-execution inconsistency introduced by retaining non-cooperative agents as service resources in the rollout model. The experimental evaluation therefore tests whether this improved alignment is sufficient to recover stable routing and rollout's empirical performance advantage over IA-RA across different adversarial and monitoring conditions.

\section{Case Study and Empirical Evaluation}
\label{sec:experiments}
We evaluate the proposed trust-aware planning framework in an online multi-robot pickup-and-delivery case study driven by real San Francisco taxicab demand data~\cite{piorkowski2009crawdad}. The experiments are organized to test the mechanisms developed in the preceding sections. We first characterize the routing impact and detectability of monitor-aware localization spoofing in the absence of a defense. We then evaluate the complementary roles of localization and behavioral trust in identifying adversarial agents. Finally, we study whether filtering the agents considered for planning using these trust estimates reduces planner-execution mismatch sufficiently for rollout to recover its empirical advantage over the IA-RA base policy.

\subsection{Experimental Setup and Evaluation Protocol}
\label{subsec:experimental_setup}

\subsubsection{Environment and Demand}
\begin{figure}[t]
    \centering
    \includegraphics[width=0.4\linewidth,
        height=0.4\textheight,
        keepaspectratio]{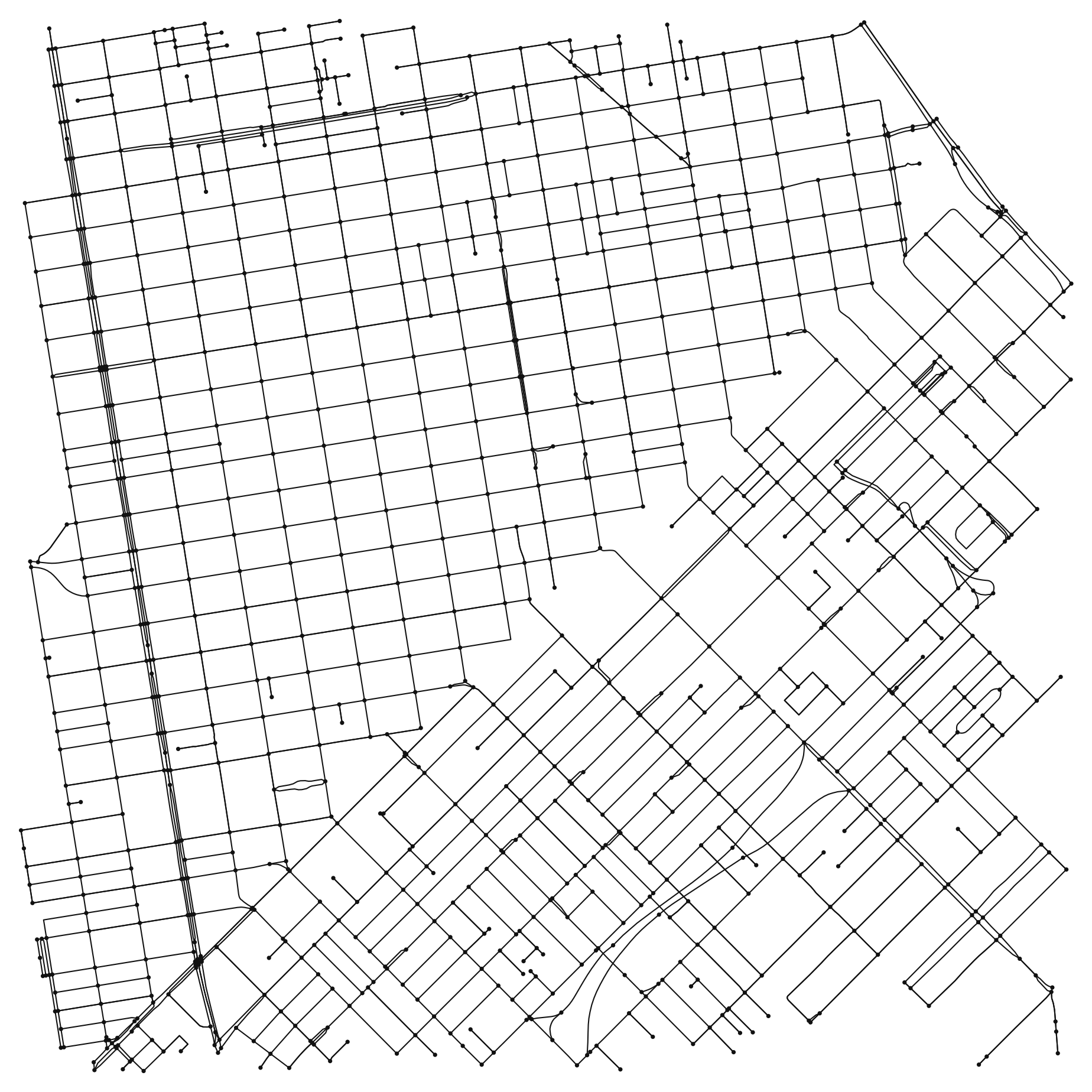}
    \vspace{-5pt}
    \caption{San Francisco road network used for the pickup-and-delivery case study. The environment contains $1026$ intersections and $2300$ directed road segments within a $1500$~m radius region.}
    \vspace{-10pt}
    \label{fig:sf_map}
\end{figure}

The experiments use a directed road network covering a $1500$~m radius region centered on San Francisco's financial district, shown in Fig.~\ref{fig:sf_map}. The resulting graph contains $1026$ intersections and $2300$ directed road segments. Each simulation time step corresponds to one minute.

Transportation demand is generated from the San Francisco taxicab dataset~\cite{piorkowski2009crawdad}. Following the demand-estimation procedure of~\cite{garces2024approximate}, the request-arrival, pickup-location, and drop-off-location distributions $p_{\eta}$, $p_{\rho}$, and $p_{\delta}$ are estimated using empirical relative frequencies. Unless otherwise stated, each experiment runs for $2880$ time steps, corresponding to two simulated days, and reported results are averaged over $100$ independently sampled demand realizations.

\subsubsection{Fleet, Adversary, and Routing Configurations}
Unless otherwise stated, the cooperative fleet contains $|\mathcal{C}_0|=35$ agents, matching the stable cooperative fleet configuration considered in~\cite{garces2024approximate}. The number of adversarial agents is varied across experiments and is reported either as $|\mathcal{A}_0|$ or through the initial adversarial fraction,
\begin{equation}
    F_0
    =
    \frac{|\mathcal{A}_0|}
    {|\mathcal{C}_0|+|\mathcal{A}_0|}
\end{equation}
Adversarial agents follow the monitor-aware spoofing strategy of Section~\ref{sec:adversarial_spoofing}. Unless otherwise stated, we consider spoofing bounds
\begin{equation}
d_{\max}
\in
\left\{
0.1D(\mathcal{G}),
0.5D(\mathcal{G})
\right\}
\end{equation}
representing relatively restrictive and permissive localization deviations. Adversarial agents coordinate their reports through the tiered matching strategy and do not service requests assigned to them.

The experiments compare four routing configurations as needed:
\begin{enumerate}
    \item \emph{IA-RA without monitoring}: adversarial agents remain among those considered for planning throughout the simulation.

    \item \emph{Rollout without monitoring}: rollout uses IA-RA as its base policy, but adversarial agents remain represented as cooperative service resources during lookahead.

    \item \emph{IA-RA with monitoring}: detected adversarial agents are removed before IA-RA computes subsequent assignments.

    \item \emph{Trust-aware rollout}: detected adversarial agents are removed before one-at-a-time rollout is applied to the resulting trusted planning state.
\end{enumerate}

Undetected adversarial agents continue to spoof their reported locations and may ignore routing commands during physical execution.

\subsubsection{Monitor and Rollout Parameters}
Localization thresholds are,
\begin{equation}
    \kappa^{\mathrm{loc}}_{\mathrm{adv}}=0.6729,
    \qquad
    \kappa^{\mathrm{loc}}_{\mathrm{coop}}=0.7629
\end{equation}
The adversarial threshold corresponds to the $25$th percentile of the nominal cooperative localization-trust distribution estimated from GPS observations in~\cite{abrar2024gps}, whereas the cooperative threshold is its mean. Unless otherwise stated, the localization window contains $W_{\mathrm{loc}}=67$ time steps. Behavioral trust is initialized using,
\begin{equation}
    \alpha_i^{\mathrm{beh}}(0)
    =
    \beta_i^{\mathrm{beh}}(0)
    =
    1
\end{equation}
corresponding to a uniform prior. We use equal behavioral evidence weights, $w_{\mathrm{pick}}
    =
    w_{\mathrm{exp}}
    =
    w_{\mathrm{churn}}
    =
    0.25$. The behavioral thresholds are, 
    $\theta^{\mathrm{beh}}_{\mathrm{adv}}=0.1, \hspace{1mm}
    \theta^{\mathrm{beh}}_{\mathrm{coop}}=0.6$
with posterior-confidence threshold $ \gamma^{\mathrm{beh}}_{\min}=0.6$.

For rollout, the nominal configuration uses a horizon $H=10$ with $M=100$ Monte Carlo demand scenarios. We additionally evaluate $(H,M)=(20,200)$ and $(H,M)=(30,300)$ to study the effect of planning depth. Candidate actions at a given decision epoch are evaluated using the same sampled demand scenarios.

\subsubsection{Evaluation Metrics}
We report both monitoring and routing performance. Monitoring performance is measured by the number of detected adversaries, the number of adversaries remaining active, cooperative-agent false positives, and adversarial detection time. Routing performance is measured using the number of outstanding requests, cumulative canceled requests, and the stage cost
defined in Section~\ref{sec:problem_formulation}. 

\subsection{Empirical Validation of Localization-Trust Evidence}
\label{subsec:empirical_characterization}
We first examine whether real GPS spoofing observations provide statistically useful localization-integrity evidence. This experiment supports the localization-trust branch of Section~\ref{subsec:localization_trust} independently of the subsequent routing experiments.

We use two real-world GPS spoofing datasets collected from aerial and ground robotic platforms. The first contains approximately $158{,}000$ nominal and spoofed observations from an autonomous aerial vehicle under attacks with different levels of stealth~\cite{8x3h-2817-22}. The second contains approximately $62{,}000$ nominal observations and $6{,}900$ adversarial observations collected from autonomous ground vehicles under multiple environmental and spoofing conditions~\cite{abrar2024gps}.

\begin{figure}[t]
    \centering
    \includegraphics[width=0.7\linewidth]
    {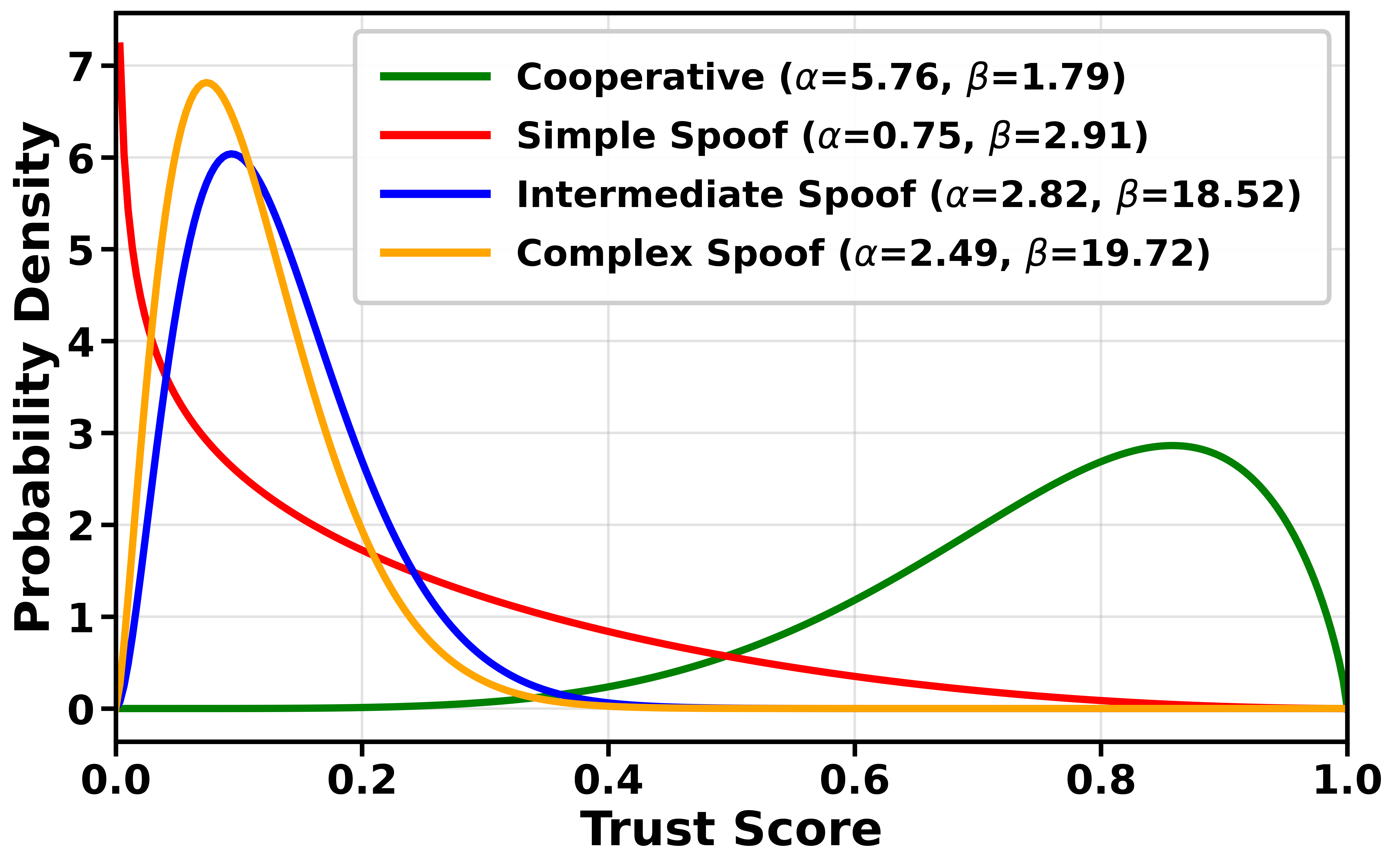}
    \vspace{-5pt}
    \caption{Empirical localization-trust distributions obtained from real GPS spoofing data. Cooperative observations concentrate at higher trust values, while spoofed observations shift toward lower trust. Increasingly stealthy attacks reduce this separation, motivating the use of complementary behavioral evidence.}
    \vspace{-10pt}
    \label{fig:signal_based_beta_distributions}
\end{figure}

Following the calibration procedure of Section~\ref{subsubsec:localization_calibration}, we use signal-level features such as pseudorange consistency, Doppler measurements, carrier-to-noise ratio, and carrier-phase continuity together with localization-quality indicators such as satellite availability, dilution of precision, and navigation error. These features produce distinct confidence distributions for cooperative and spoofed operation as shown in Fig.~\ref{fig:signal_based_beta_distributions}.

The resulting distributions show that localization signals provide useful evidence of spoofing, but also illustrate a limitation of relying on localization evidence alone: increasingly stealthy attacks move closer to the cooperative distribution. This motivates the monitor-aware attacks studied next and the behavioral evidence branch evaluated later.

\subsection{Routing Impact and Detectability of Monitor-Aware Spoofing}
\label{subsec:spoofing_without_monitoring}
We next remove the monitor entirely in order to isolate the effect of the adversarial strategy. IA-RA is used throughout these experiments, and all adversarial agents remain active for the full simulation.

\begin{figure}[t]
    \centering
    \includegraphics[width=0.99\linewidth]
    {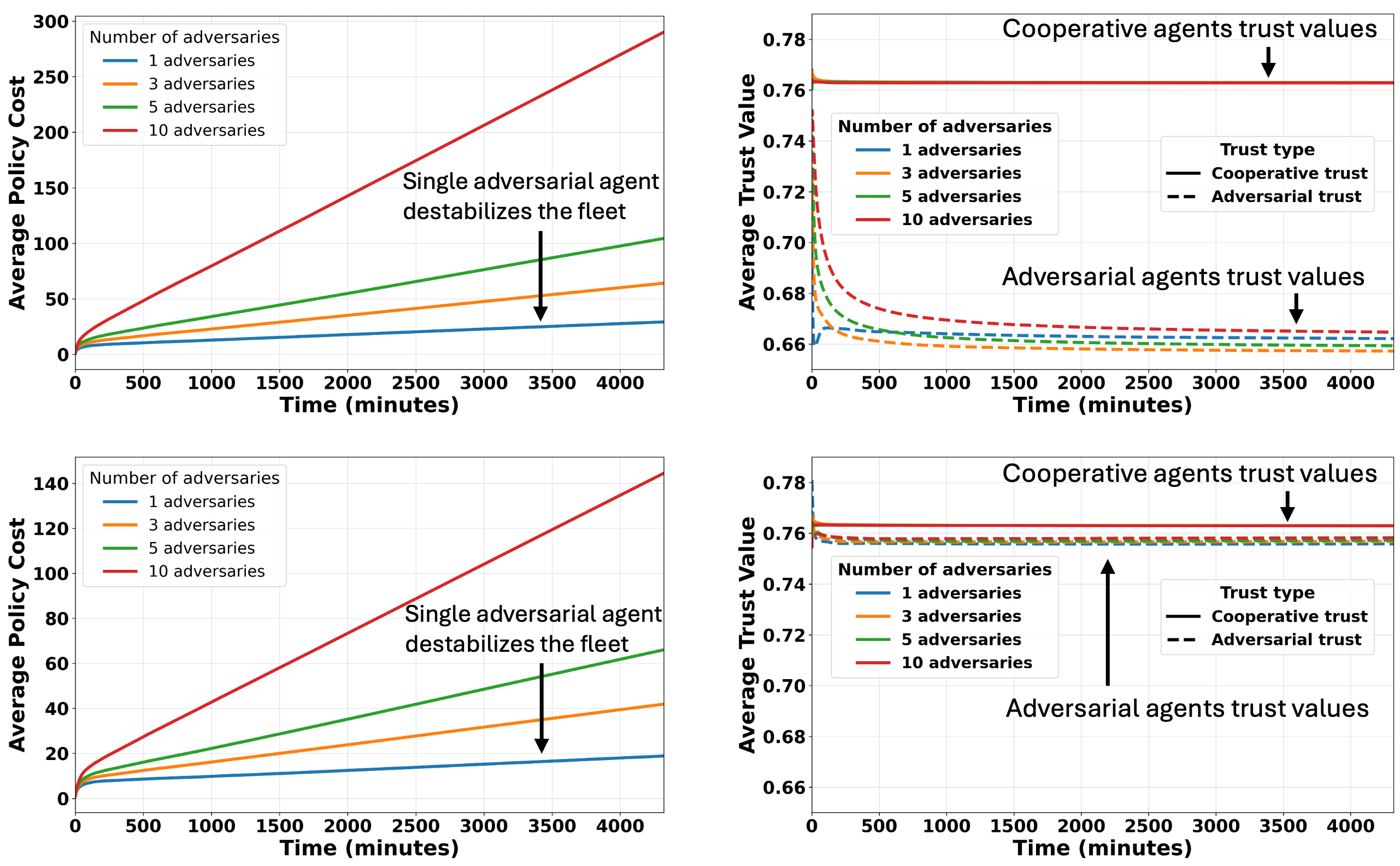}
    \vspace{-5pt}
    \caption{Policy cost and localization trust under unmonitored spoofing. Rows correspond to different maximal spoofing ranges distance-constrained
    attacks, while the left and right columns report policy cost and average team localization trust, respectively.}
    \label{fig:constrained_spoofing_model}
    \vspace{-10pt}
\end{figure}

Figure~\ref{fig:constrained_spoofing_model} compares policy cost and localization trust under different adversarial fleet sizes and spoofing bounds. Across the evaluated attack configurations, the policy cost exhibits sustained growth over the simulation horizon, consistent with loss of stability under the criterion of Section~\ref{sec:problem_formulation}. 
\begin{remark}
\label{remark:single_adversary_instability}
\textbf{A Single Adversarial Agent Can Destabilize a Nominally Stable Fleet:}
{\normalfont
The experiments show that a single adversarial agent is sufficient to destabilize a fleet operating under IA-RA whose number of cooperative agents satisfies the stability condition under fully cooperative operation. This occurs under both unconstrained and sufficiently permissive distance-constrained spoofing, demonstrating that even limited adversarial presence can induce system-level routing instability.}
\end{remark}

Restricting the spoofing radius does not eliminate the routing-level effect. As shown in Fig.~\ref{fig:constrained_spoofing_model}, the tiered strategy continues to identify requests through which adversarial agents can interfere with cooperative service and induce persistent degradation.

The corresponding localization-trust trajectories expose the influence-detectability trade-off. Cooperative-agent trust remains approximately unchanged because cooperative agents report their locations truthfully. Adversarial trust decreases under spoofing, but the separation from cooperative trust becomes smaller when the maximum spoofing distance is reduced. Thus, distance-constrained adversaries can continue to degrade routing while producing localization evidence that is more difficult to distinguish from nominal operation.

These results provide the first link in the experimental narrative: localization spoofing creates substantial planner-execution mismatch, while monitor-aware attackers can reduce the localization evidence available for detecting that mismatch.

\subsection{Complementary Value of Localization and Behavioral Evidence}
\label{subsec:monitor_ablation}
We next evaluate whether the two evidence branches introduced in Section~\ref{sec:trust_monitor} provide complementary information. We compare a localization-only monitor, a behavioral-only monitor, and the full fused monitor.

\begin{figure}[t]
    \centering
    \includegraphics[width=0.99\linewidth]
    {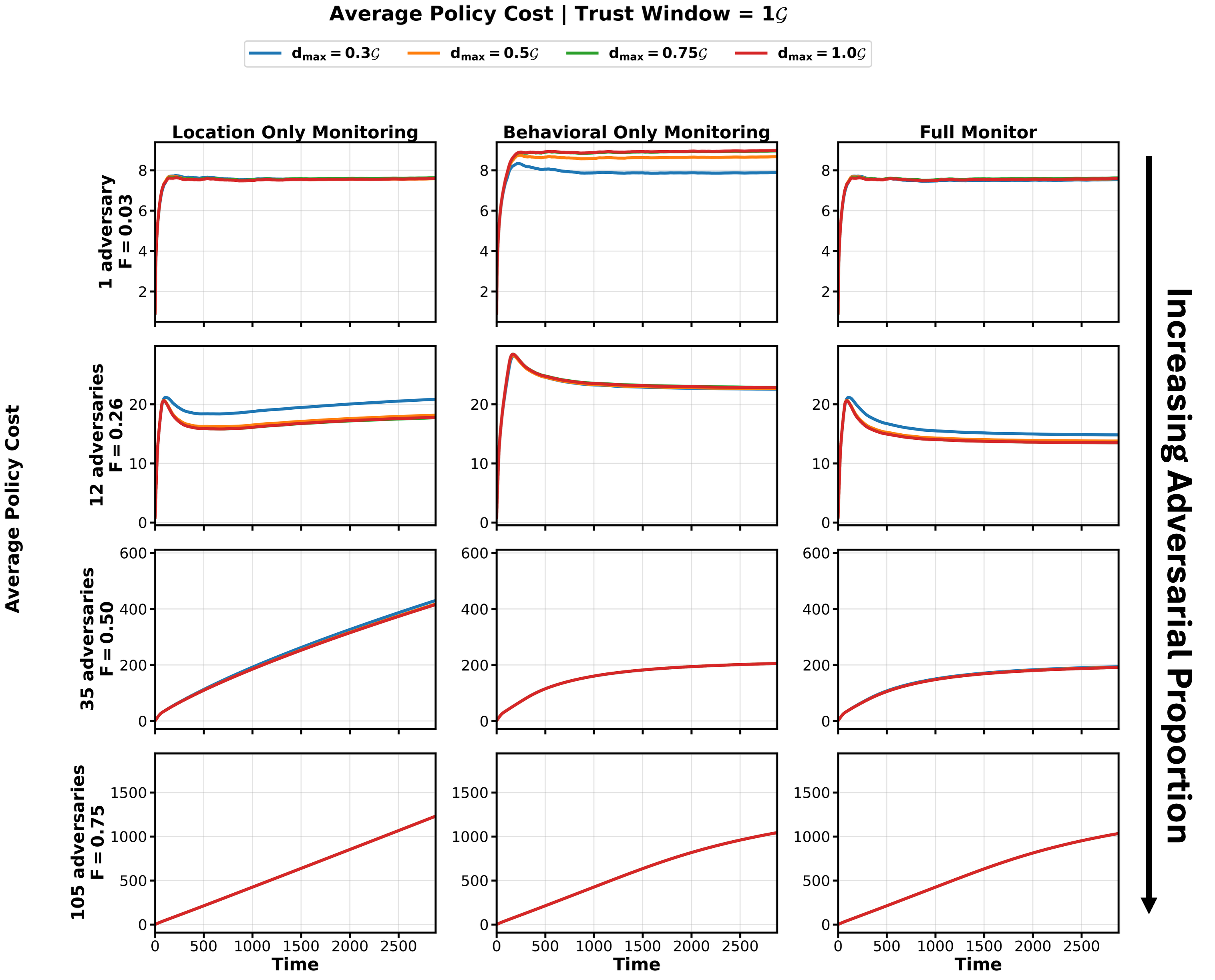}
    \vspace{-5pt}
    \caption{Average routing cost for the localization-only, behavioral-only, and fused monitors. Rows vary the adversarial fleet fraction, columns vary the monitoring configuration, and curves vary the spoofing distance.}
    \vspace{-7pt}
    \label{fig:monitor_ablation_study_policy_cost}
\end{figure}

\begin{figure}[t]
    \centering
    \includegraphics[width=0.99\linewidth]
    {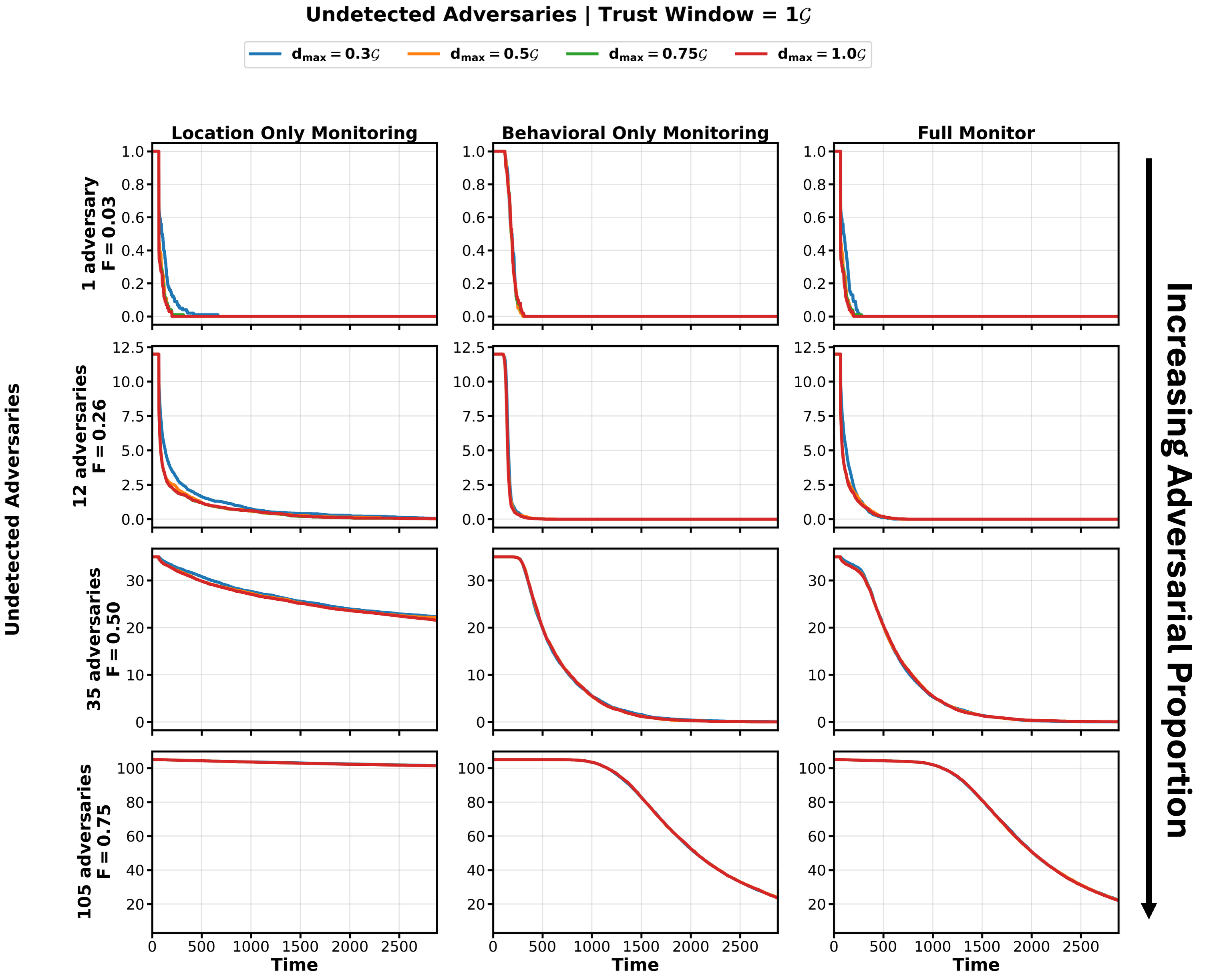}
    \vspace{-5pt}
    \caption{Average number of adversarial agents remaining active under the monitor-ablation configurations. The layout matches Fig.~\ref{fig:monitor_ablation_study_policy_cost}.}
    \vspace{-10pt}
    \label{fig:monitor_ablation_study_undetected_advs}
\end{figure}

\begin{figure}[t]
    \centering
    \includegraphics[width=0.99\linewidth]
    {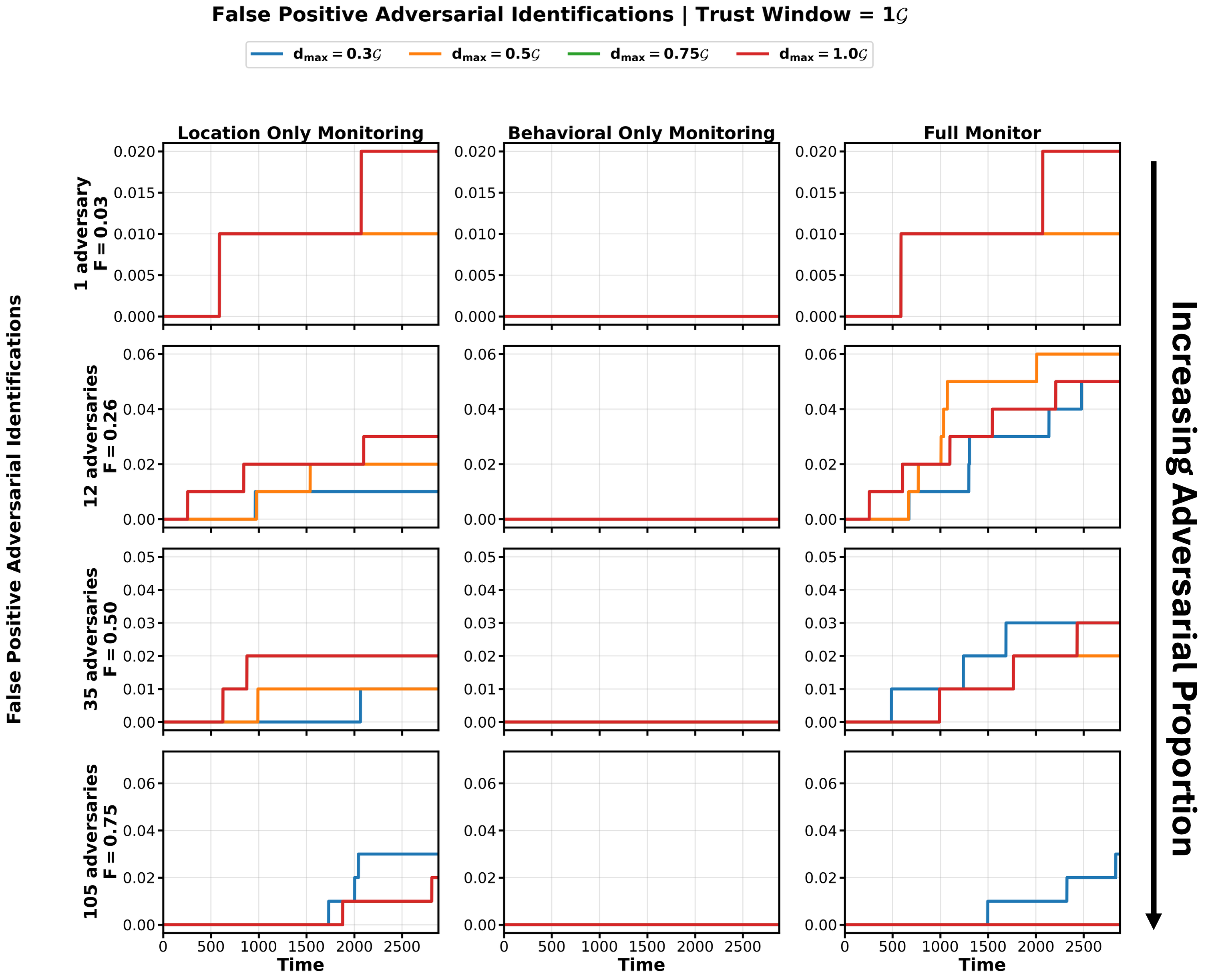}
    \vspace{-5pt}
    \caption{Average number of cooperative agents incorrectly removed by the monitor.}
    \vspace{-7pt}
    \label{fig:monitor_ablation_study_false_positives}
\end{figure}

The localization branch responds directly to inconsistencies in reported position information and can therefore identify sufficiently large spoofing deviations relatively early. Its effectiveness decreases, however, as the attacker reduces the localization discrepancy. The behavioral branch does not depend directly on spoofing magnitude. Instead, it accumulates evidence from successful pickups, expirations, and assignment churn. Behavioral detection can therefore require more time, but it remains informative when the localization signal is ambiguous. Figures~\ref{fig:monitor_ablation_study_policy_cost}-\ref{fig:monitor_ablation_study_false_positives} show the resulting trade-offs.

The results confirm the complementary roles of the two evidence sources. Localization evidence identifies some adversaries earlier, whereas behavioral evidence ultimately identifies adversaries whose localization behavior remains less distinguishable from cooperative operation. The fused monitor yields the lowest routing cost and the fewest adversarial agents remaining active across the evaluated configurations while maintaining a low cooperative-agent false-positive rate. The behavior-only monitor produces no cooperative false positives in the reported ablation.

\subsection{Trust-Aware Rollout Under Planner-Execution Mismatch}
\label{subsec:trust_monitor_rollout_experiments}
We now evaluate the complete interaction between adversarial spoofing, online monitoring, and rollout planning. The objective is not to assume that monitoring automatically guarantees rollout improvement, but to test whether removing unreliable agents reduces planner-execution mismatch sufficiently for rollout to recover its empirical benefit.

We compare IA-RA and rollout both with and without monitoring. The key comparison is between rollout without monitoring and trust-aware rollout. Without monitoring, adversarial agents remain represented as cooperative service resources in the rollout model even though they may not execute the corresponding actions. With monitoring, detected adversaries are removed from the trusted planning state before rollout is performed.

\subsubsection{Adversarial Removal and Planning-Execution Alignment}
\begin{figure}[t]
    \centering
    \includegraphics[width=0.9\linewidth]{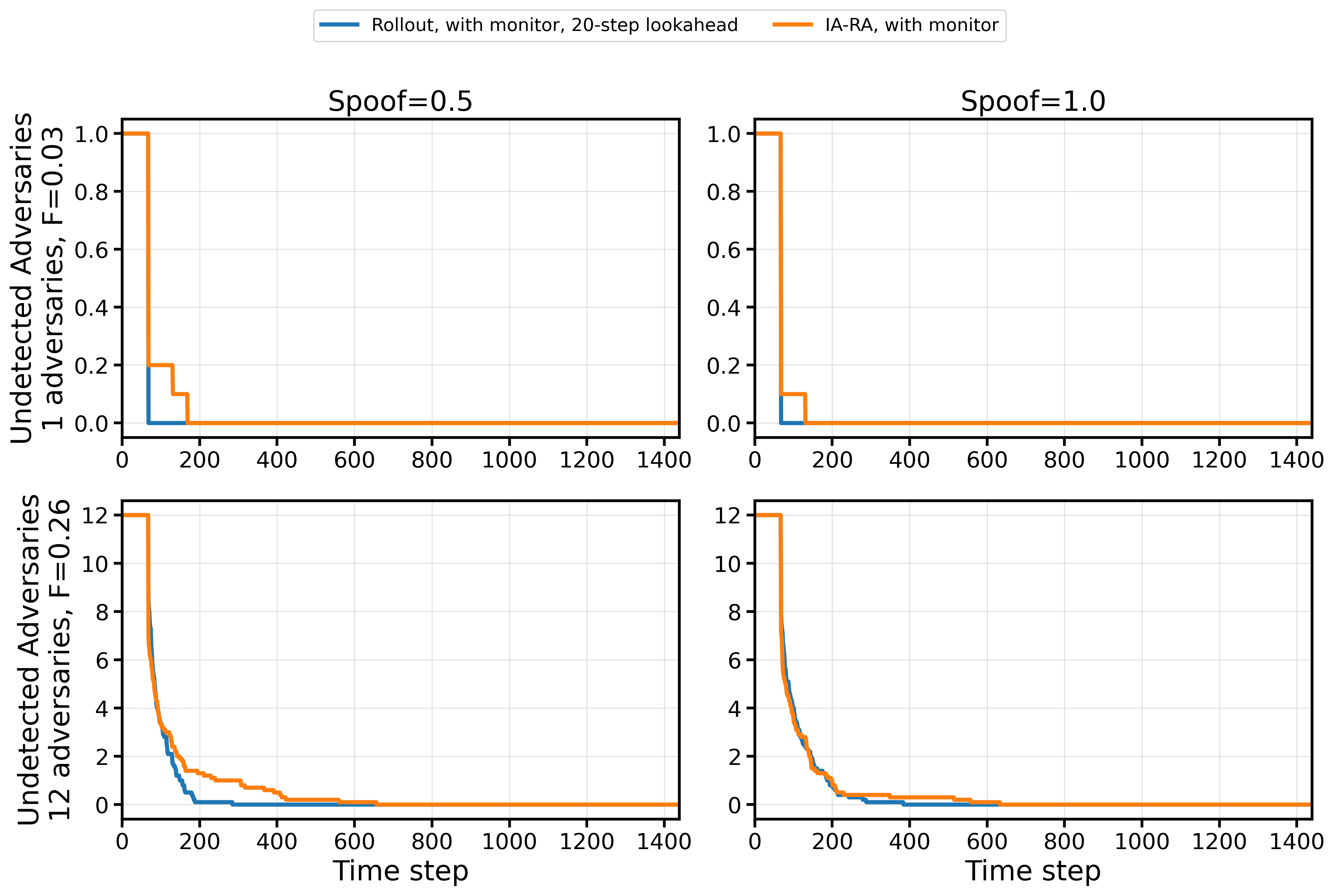}
    \vspace{-7pt}
    \caption{Average number of adversarial agents remaining active for the IA-RA and rollout configurations. Rows vary the initial adversarial fleet size and columns vary the spoofing radius. Monitoring progressively removes adversarial agents from consideration during planning.}
    \vspace{-10pt}
    \label{fig:trust_aware_undetected_advs}
\end{figure}

Figure~\ref{fig:trust_aware_undetected_advs} first isolates the mechanism through which monitoring changes the planning problem. In the unmonitored configurations, all adversarial agents remain active throughout the simulation. In the monitored configurations, the number of active adversaries decreases as trust evidence accumulates and agents are removed. No cooperative false positives occur in the configurations shown in this experiment.

This reduction in the number of active adversarial agents is the mechanism by which trust changes rollout: the lookahead model is progressively restricted to agents that are more consistent with the cooperative execution model.

\subsubsection{Outstanding Demand}
\begin{figure}[t]
    \centering
    \includegraphics[width=0.9\linewidth]{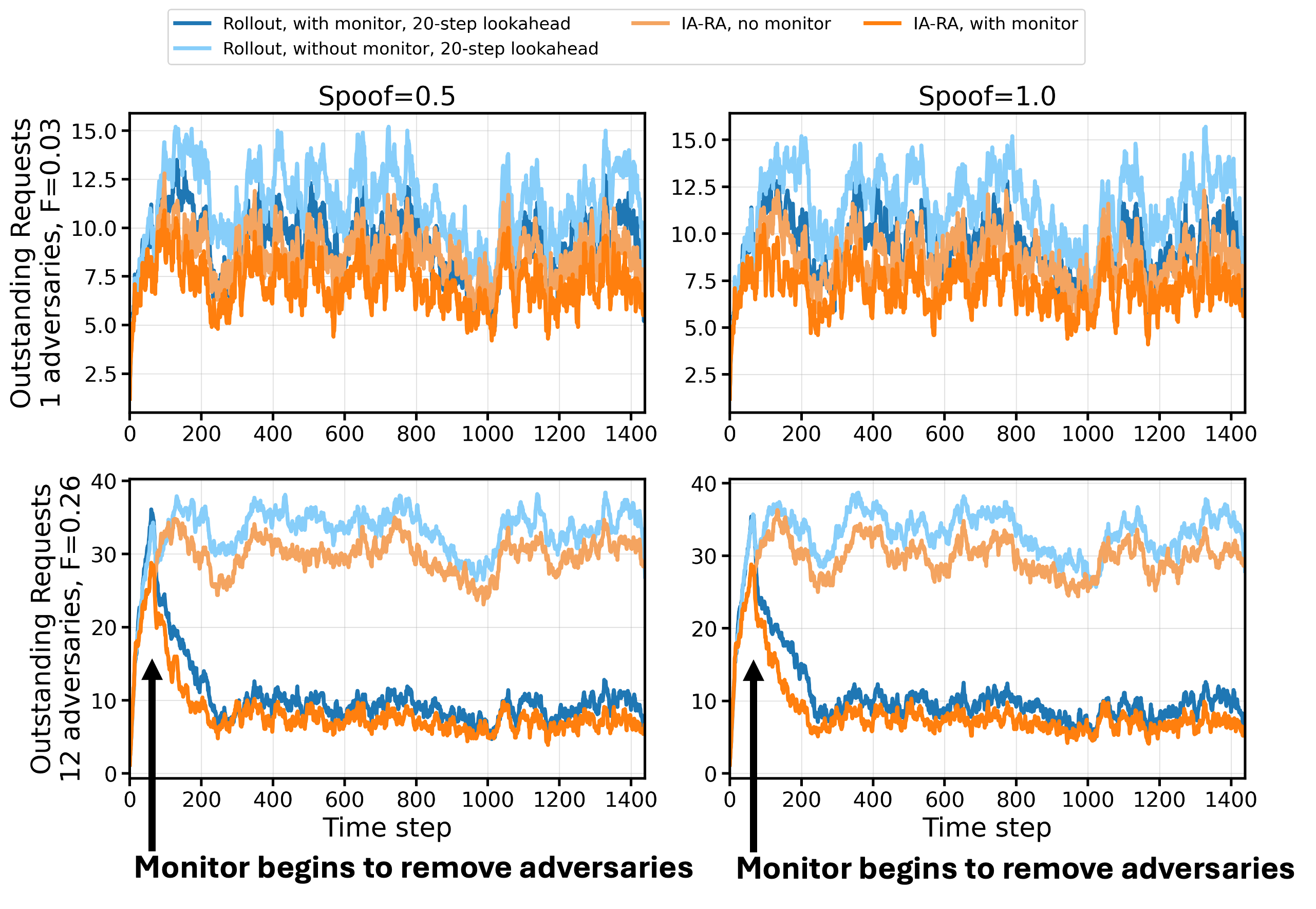}
    \vspace{-7pt}
    \caption{Average number of outstanding requests for the IA-RA and rollout configurations. Adversarial removal reduces the mismatch between simulated and executed service, allowing monitored policies to reduce the accumulated backlog. For sufficiently long lookahead, trust-aware rollout approaches or improves upon the backlog of monitored IA-RA.}
    \label{fig:trust_aware_outstanding_reqs}
    \vspace{-7pt}
\end{figure}

Figure~\ref{fig:trust_aware_outstanding_reqs} shows the corresponding effect on outstanding requests. Rollout without monitoring can maintain a large backlog because simulated trajectories assign service capability to agents that do not provide that service during physical execution.

After adversarial removal, the monitored configurations reduce this backlog. Trust-aware rollout eventually reaches outstanding-request levels comparable to those obtained by monitored IA-RA. Outstanding request levels can be improved by considering longer lookahead lengths and bigger fleet sizes as shown in the following subsections. This indicates recovery in current service responsiveness after adversarial agents have been removed from planning.

\subsubsection{Cancellations and Transient Mismatch}
Outstanding requests capture the current backlog, whereas cancellations retain the effect of earlier service failures. This distinction is important for interpreting trust-aware rollout.

\begin{figure}[t]
    \centering
    \includegraphics[width=0.9\linewidth]{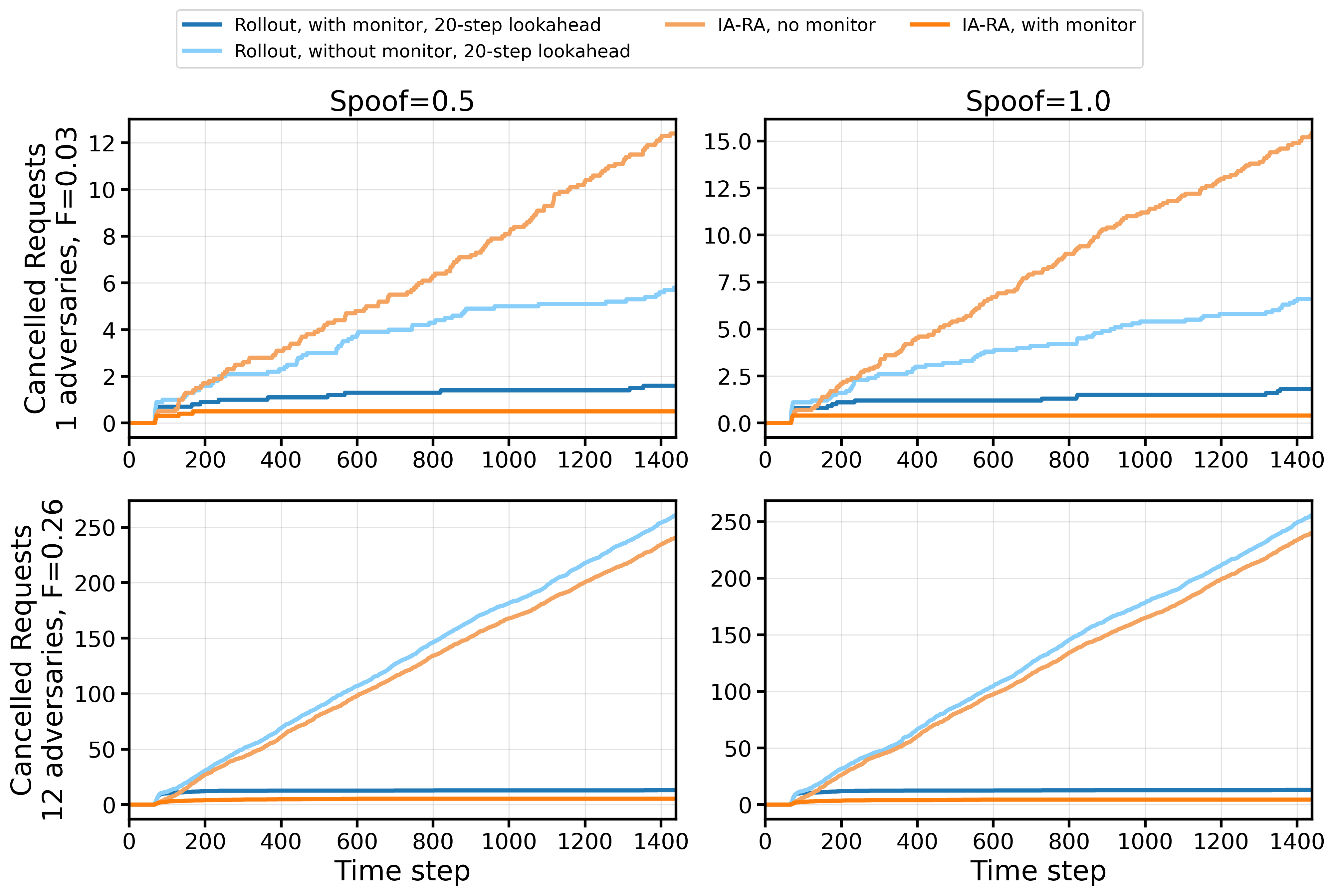}
    \vspace{-7pt}
    \caption{Average cumulative canceled requests. Without monitoring, service failures continue to accumulate because adversarial agents remain represented as cooperative resources. Monitoring can arrest this growth after adversarial removal.}
    \label{fig:trust_aware_cancelled_reqs}
\end{figure}

Figure~\ref{fig:trust_aware_cancelled_reqs} shows that rollout without monitoring continues to accumulate cancellations because adversarial agents repeatedly invalidate the service trajectories represented during lookahead. Once monitoring removes those agents, cancellations can stabilize. The effect, however, depends strongly on the rollout horizon as we will show in the following subsection.

\subsubsection{Routing Cost and Recovery}
\begin{figure}[t]
    \centering
    \includegraphics[width=0.9\linewidth]{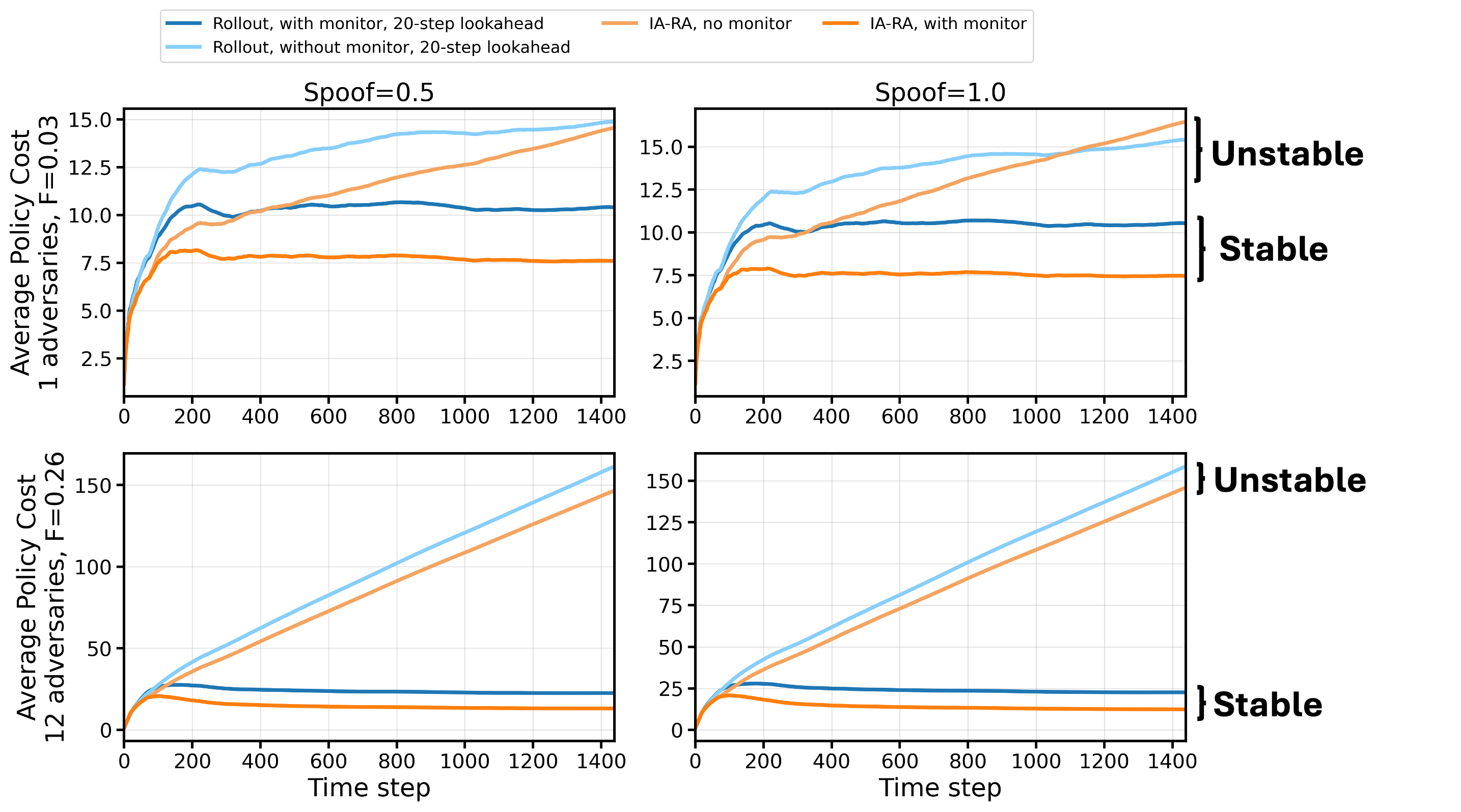}
    \vspace{-7pt}
    \caption{Average routing policy cost for the IA-RA and rollout configurations. The cost combines current backlog and cumulative cancellations. Trust-aware rollout may therefore retain a larger cost after detection even when its current outstanding-request count has recovered, because early cancellations remain in the cumulative term.}
    \label{fig:trust_aware_policy_cost}
\end{figure}

The stage-cost results in Fig.~\ref{fig:trust_aware_policy_cost} should be interpreted jointly with the backlog and cancellation trajectories. Because
\begin{equation}
    g_t
    =
    |\bar{\mathcal{R}}_t|
    +
    |\mathcal{R}^{\mathrm{can}}_{0:t}|
\end{equation}
cancellations accumulated before adversarial removal remain part of the cost for the remainder of the simulation.

This explains why trust-aware rollout can exhibit a higher cost than monitored IA-RA in some configurations even after its current backlog becomes comparable to or smaller than that of the base policy. The difference reflects service failures accumulated during the initial period in which adversarial agents remained among those considered for planning.

Taken together, these results distinguish two regimes. Before detection, rollout can be particularly vulnerable because lookahead compounds the consequences of an incorrect execution model. After adversarial removal, the source of mismatch associated with adversarial non-execution is reduced and rollout can again exploit lookahead to improve current service decisions. Whether that recovery is sufficient to overcome the transient damage depends on the planning horizon and available cooperative fleet capacity.

\subsection{Effect of Planning Horizon and Cooperative Capacity}
\label{subsec:rollout_horizon_capacity}

The rollout experiments indicate that recovery depends not only on adversarial detection but also on the planner's ability to compensate for the backlog accumulated before detection. We therefore examine two factors that affect this recovery: rollout horizon and cooperative fleet capacity.

\begin{figure}[t]
    \centering
    \includegraphics[width=0.9\linewidth]{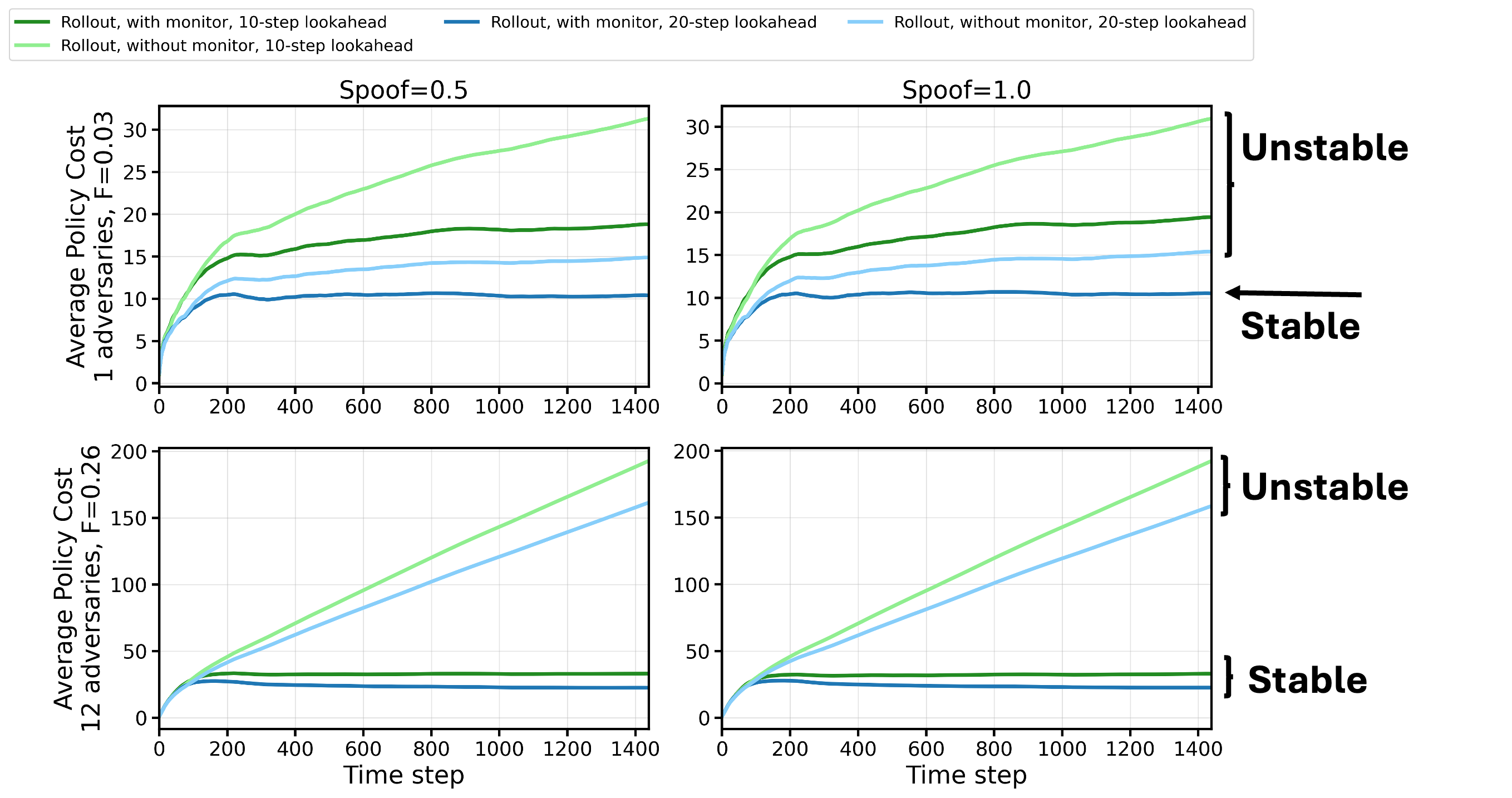}
    \vspace{-7pt}
    \caption{Average routing policy cost for two rollout configurations with different lookahead lengths of $H=10$ and $H=20$. The plot shows that the short-horizon case can remain unstable even with monitoring, whereas longer horizons allow the planner to better account for the downstream consequences of current assignments and lead to stable policies. }
\label{fig:trust_aware_different_rollout_lookaheads}
\end{figure}

Increasing the rollout horizon from $H=10$ to longer lookahead values reduces persistent backlog and cancellations in the evaluated configurations. The short-horizon case can remain unstable even with monitoring, whereas longer horizons allow the planner to better account for the downstream consequences of current assignments. We therefore interpret trust-based removal as restoring a more reliable set of agents for planning, while the rollout horizon determines how effectively the planner can use that corrected representation to recover from prior service degradation. These effects can be observed in Fig. \ref{fig:trust_aware_different_rollout_lookaheads} that compares the policy cost for rollout with $H=10$ and $H=20$ lookaheads.

For $H=10$, cancellations can continue to accumulate even after monitoring is introduced. Thus, adversarial removal alone does not erase the transient damage accumulated while planner-execution mismatch was present, nor does it guarantee that a short-horizon rollout will recover. Longer lookahead horizons reduce this effect and allow cancellations to stabilize in the evaluated configurations.

\begin{figure}[t]
    \centering
    \includegraphics[width=0.9\linewidth]
    {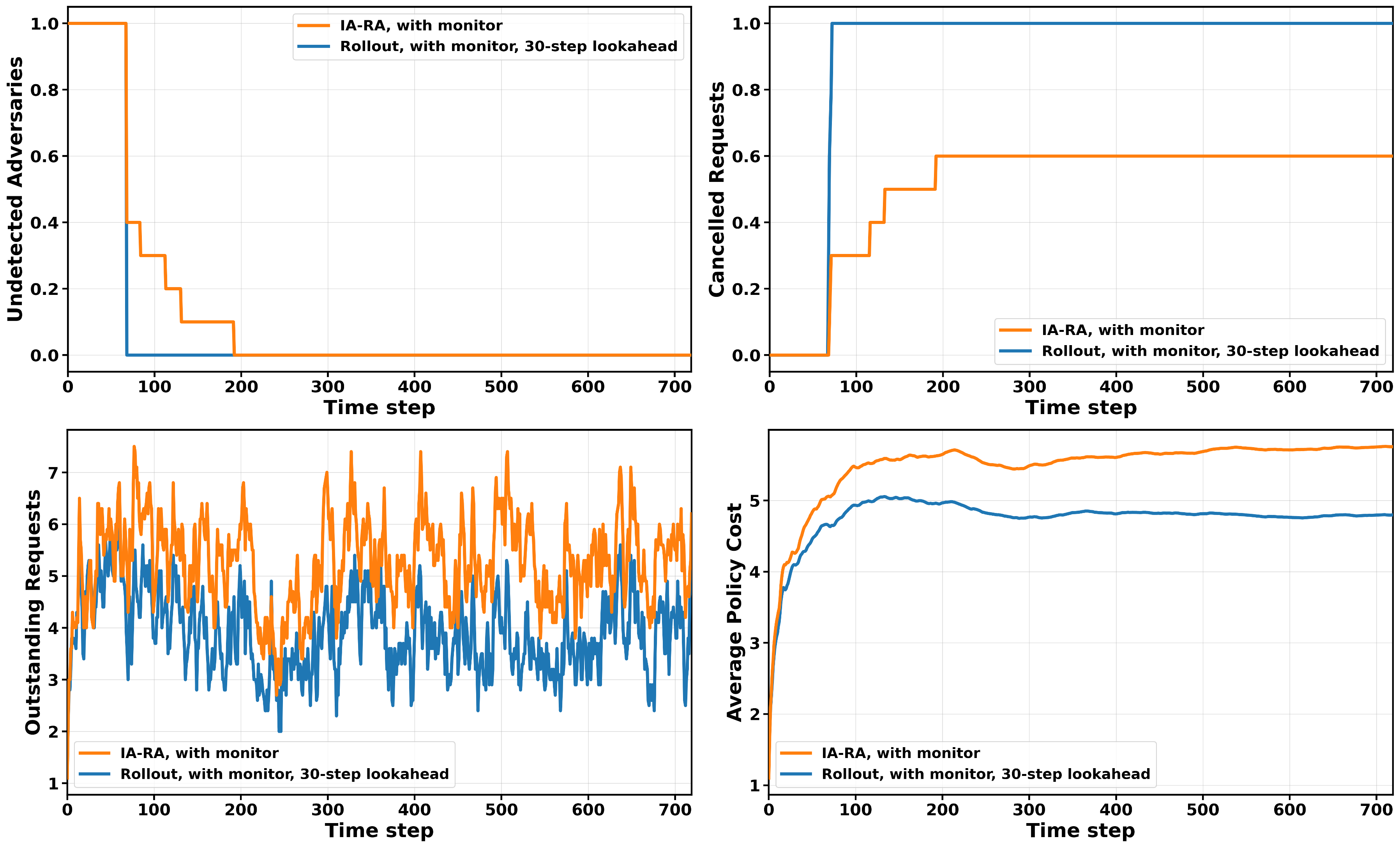}
    \vspace{-7pt}
    \caption{Monitored IA-RA and trust-aware rollout with $49$ cooperative agents, one adversarial agent, and $H=30$. The panels report adversarial agents remaining active, cumulative cancellations, outstanding requests and routing cost. With additional cooperative capacity, trust-aware rollout achieves lower number of outstanding requests and lower policy costs than monitored IA-RA.}
    \label{fig:trust_aware_policy_cost_49coops_1_adv}
\end{figure}

The preceding experiments use $35$ cooperative agents, corresponding to the minimum cooperative fleet configuration considered in the stable baseline setting. Operating close to this capacity boundary leaves little surplus service capacity for recovering from backlog accumulated during the adversarial detection interval. To examine this effect, we repeat the trust-aware rollout experiment with $49$ cooperative agents, one adversarial agent, and rollout horizon $H=30$. The results for this new setting are depicted in Fig.~\ref{fig:trust_aware_policy_cost_49coops_1_adv}.

In this higher-capacity setting, trust-aware rollout achieves lower routing cost and fewer outstanding requests and cancellations than the monitored IA-RA base policy in the reported experiment. This result suggests that the transient planner-execution mismatch is especially costly when the cooperative fleet operates near its minimum stable capacity. Additional cooperative capacity provides greater ability to clear the backlog accumulated before adversarial removal and makes the post-detection benefit of rollout more pronounced.

\section{Discussion and Limitations}
\label{sec:discussion}
The results support the central premise of this work: trust is most useful for sequential planning when it changes the representation on which decisions are computed, rather than remaining an auxiliary detection signal. In the routing instantiation considered here, localization and behavioral evidence are used to determine which agents remain under consideration during planning. This filtering operation directly addresses one source of planner-execution mismatch by preventing agents with sufficient evidence of unreliable behavior from continuing to influence routing and rollout simulations.

At the same time, the experiments also show that trust-aware filtering does not immediately remove the consequences of prior mismatch. Adversarial agents may remain active while evidence is accumulated, and requests delayed or canceled during this interval continue to affect subsequent system performance. The effectiveness of trust-aware planning therefore depends not only on eventual detection, but also on detection delay, cooperative fleet capacity, and the ability of the planner to recover from the backlog accumulated before removal.

\subsection{Interpretation of the Empirical Results}
\label{subsec:discussion_results}
The localization and behavioral trust branches provide complementary views of reliability. Localization trust evaluates the integrity of the information entering the planner, whereas behavioral trust evaluates whether the resulting task-level execution is consistent with cooperative operation. This distinction becomes particularly important under monitor-aware spoofing. Reducing localization deviation can make an adversary more difficult to distinguish from a cooperative agent using localization evidence alone, but it does not necessarily eliminate the downstream effects of repeatedly attracting assignments and failing to service them. The empirical ablation results therefore support the use of both information-level and execution-level evidence in constructing the trusted planning state.

The rollout experiments further illustrate why this distinction matters for model-based planning. While undetected adversaries remain in $\mathcal{L}_{t^+}$, the rollout model can represent them as cooperative resources even though their executed actions differ from the commands used during lookahead. Trust-aware removal reduces this specific source of mismatch by changing the set of agents over which rollout is performed.

This mechanism should not be interpreted as an unconditional restoration of rollout performance. Detection itself is delayed, finite-horizon rollout remains approximate, and the system may accumulate substantial backlog before the set of agents considered for planning is corrected. In particular, the experiments show that a short rollout horizon can remain insufficient to recover from the transient degradation produced before adversarial removal. Additional cooperative capacity makes this recovery easier by providing greater ability to service accumulated demand after the mismatch has been reduced. Thus, adversarial removal improves the consistency of the planning representation, while the subsequent recovery still depends on the planning horizon and available system capacity.

\subsection{Limitations}
\label{subsec:limitations}
Several limitations define the scope of the present study. First, the adversarial model intentionally provides the attackers with substantial knowledge and coordination capability, but restricts their behavior in other ways. Adversarial agents manipulate localization reports, coordinate their targets, and fail to service assigned requests, but they cannot create new identities. The experiments therefore do not cover Sybil attacks, intermittent identity changes, compromise of the central monitor, or adversaries that manipulate additional state variables. Similarly, the considered adversaries either fail to service requests or follow the cooperative model. Partially compliant adversaries that strategically mix successful and failed service to manipulate behavioral trust are an important direction for future study.

Second, localization trust requires a source of integrity evidence that is sufficiently independent of the reported localization information. In simulation, the normalized discrepancy can be generated using the true agent position. In deployment, this quantity must instead be derived from an independent localization-integrity mechanism, redundant sensing, or another trusted reference. The performance of the complete system will therefore depend on the quality, availability, and potential failure modes of this external source. The trust distributions calibrated from the GPS spoofing datasets considered in this work may also change across sensing hardware, environments, geographic regions, or previously unseen attack strategies. Recalibration or online adaptation may be required under distribution shift.

Third, behavioral trust relies on the ability to reconstruct request histories and observe terminal request outcomes with sufficient reliability. The responsibility rule used here attributes positive evidence to successful service and negative evidence to expiration and assignment churn. These signals are informative for the pickup-and-delivery problem considered in this paper, but they do not uniquely identify the cause of every service failure. Congestion, hardware faults, communication loss, or other non-adversarial disturbances may produce similar outcomes. More expressive attribution mechanisms could distinguish among these causes or explicitly represent uncertainty in behavioral responsibility.

Fourth, the enforcement mechanism used in this work permanently removes an agent once it is classified as adversarial. This provides a simple and conservative planning interface, but removal is not necessarily the appropriate response in every robotic system. False-positive removal reduces available fleet capacity and can itself degrade routing performance, particularly when the system operates close to its minimum service capacity. Alternative interventions could include temporary quarantine, trust-dependent action constraints, reduced assignment priority, redundant verification, or reversible re-entry after additional evidence is collected.

Finally, the stability and rollout results reported here are empirical. The simulations demonstrate sustained degradation under adversarial operation and recovery under several trust-aware configurations, but they do not provide a general analytical characterization of detection time, closed-loop stability, or rollout improvement under planner-execution mismatch. Classical rollout improvement arguments apply to the model used for lookahead; when the executing system differs from that model, the corresponding real-system improvement interpretation need not hold. Moreover, the Monte Carlo finite-horizon rollout used here introduces additional approximation through finite lookahead and sampled demand. Developing formal conditions relating trust-classification accuracy, detection delay, the number of remaining adversarial agents, fleet capacity, and closed-loop performance remains an important direction for future work.

\subsection{Broader Applicability}
\label{subsec:broader_applicability}
Although the experiments focus on localization spoofing in online multi-robot routing, the underlying planning principle is not specific to GPS or pickup-and-delivery systems. The framework requires two types of observable evidence: an information-level signal that provides some indication of the reliability of the state presented to the planner, and execution-level evidence that reveals whether the resulting physical behavior is consistent with the planner's assumptions. These signals are used to update a reliability assessment, which in turn modifies the state or set of agents considered in subsequent planning.

This abstraction arises in many robotic systems. Information-level evidence may be obtained from localization integrity, perception confidence, communication quality, sensor consistency, actuator diagnostics, or redundant state estimation. Behavioral evidence may be obtained from successful task completion, missed deadlines, execution failures, repeated reallocations, or other verifiable task outcomes. The appropriate trust model, attribution mechanism, and intervention would depend on the application.

The same principle is particularly relevant to model-based sequential decision-making methods. Rollout, model predictive control, approximate dynamic programming, and other receding-horizon approaches evaluate actions using an internal representation of the system expected to execute those actions. When unreliable agents, sensors, communication channels, or actuators create persistent disagreement between that representation and physical execution, planning quality can degrade even if the optimization method itself is unchanged. Trust-aware state construction provides one mechanism for adapting the planning representation as reliability evidence accumulates.

In the routing instantiation studied here, that adaptation is implemented by removing agents classified as adversarial from $\mathcal{L}_{t^+}$. More generally, trust need not result in binary removal. It may instead modify resource availability, constrain feasible actions, alter the weight assigned to uncertain information, or trigger a fallback planning model. The broader contribution of the framework is therefore the interface between reliability assessment and sequential planning: observable evidence is used to determine which information and resources should be represented when future decisions are computed.

\section{Conclusion}
\label{sec:conclusion}

This paper studies resilient sequential planning in multi-robot systems when agents can both corrupt the information presented to the planner and deviate from the actions assumed during planning. We show that these two effects create a planner-execution mismatch: the planning state can represent agents and service capabilities that are not realized by the physical system. To study this mismatch under strategic interference, we introduced a monitor-aware localization-spoofing model in which adversarial agents coordinate their reports while limiting spoofing magnitude, thereby trading routing influence against detectability.

We address this problem through a trust-aware planning architecture that uses observable reliability evidence to construct the state on which subsequent decisions are made. In the routing instantiation, localization-integrity evidence and request-level behavioral outcomes are combined to classify agents and remove those with sufficient evidence of adversarial behavior from the agents considered during planning. This makes trust operational within the decision loop: rather than serving only as a detection score, trust determines which agents are considered during planning.

The empirical results demonstrate the consequence of this coupling for rollout-based routing. Undetected adversaries can cause rollout simulations to diverge from physical execution and can eliminate the empirical advantage of lookahead over the IA-RA base policy. As adversarial agents are identified and removed, the set of agents considered during planning becomes better aligned with the executing fleet, allowing rollout to recover when sufficient planning horizon and cooperative capacity are available. These results support a broader principle for resilient sequential decision making: when planning performance depends on an internal model of the executing system, reliability estimates are most useful when they are used to adapt that model and the state over which future decisions are computed.

\section{Acknowledgements}
This work was supported in part by the Defense Advanced Research Projects Agency (DARPA) under Grant No. D24AP00319-00. The views and conclusions expressed in this paper are those of the authors and do not reflect the official policy or position of the U.S. Army, U.S. Department of War, or U.S. Government.

\bibliographystyle{IEEEtran}
\bibliography{refs}

\end{document}